\documentclass[aps, prb, twocolumn, showkeys, superscriptaddress, nofootinbib, floatfix, longbibliography]{revtex4-2}

\usepackage[utf8]{inputenc}
\usepackage[T1]{fontenc}

\usepackage[english]{babel}
\usepackage{amsmath}
\usepackage{amssymb}
\usepackage{array}
\usepackage{booktabs}

\usepackage[pdftex]{graphicx}
\usepackage[hypcap=true]{caption}
\usepackage[hypcap=true, labelformat=simple]{subcaption}
\usepackage{ragged2e}
\DeclareCaptionJustification{justified}{\justifying}
\usepackage{color}
\usepackage[usenames, dvipsnames]{xcolor}
\usepackage[linktoc=page, colorlinks=true, linkcolor=MidnightBlue, citecolor=MidnightBlue, urlcolor=OliveGreen]{hyperref}

\newcommand{\F}{\mathrm{\scriptscriptstyle F}}

\newcommand{\A}{{\mathrm{\scriptscriptstyle A}}}
\renewcommand{\L}{{\mathrm{\scriptscriptstyle L}}}
\newcommand{\R}{{\mathrm{\scriptscriptstyle R}}}
\newcommand{\C}{{\mathrm{\scriptscriptstyle C}}}
\newcommand{\LL}{{\mathrm{\scriptscriptstyle LL}}}
\newcommand{\LR}{{\mathrm{\scriptscriptstyle LR}}}
\newcommand{\RL}{{\mathrm{\scriptscriptstyle RL}}}
\newcommand{\RR}{{\mathrm{\scriptscriptstyle RR}}}

\newcommand{\muB}{\mu_\mathrm{\scriptscriptstyle B}}
\newcommand{\ESO}{E_\mathrm{\scriptscriptstyle SO}}

\newcommand{\EF}{E_{\F}}
\newcommand{\kF}{k_{\F}}
\newcommand{\kpar}{\mathbf{k}_{\parallel}}
\newcommand{\vF}{v_{\F}}
\newcommand{\BZ}{{\mathrm{\scriptscriptstyle BZ}}}

\newcommand{\Vex}{V_\mathrm{ex}}
\newcommand{\lMFP}{l_\mathrm{\scriptscriptstyle MFP}}

\newcommand{\Jc}{J_\mathrm{c}}
\newcommand{\Ic}{I_\mathrm{c}}

\newcommand{\G}{G}
\newcommand{\Gsigma}{G_{\sigma}}
\newcommand{\Gup}{G_\uparrow}
\newcommand{\Gdn}{G_\downarrow}

\newcommand{\GF}{\mathcal{G}^\mathrm{r}}

\begin{document}

\title{Supercurrent decay in ballistic magnetic Josephson junctions}

\author{Herv\'e Ness}
\affiliation{Department of Physics, King's College London, Strand Campus, London WC2R 2LS, UK}

\author{Ivan A. Sadovskyy}
\affiliation{Microsoft Quantum, Microsoft Station Q, University of California, Santa Barbara, California 93106, USA}

\author{Andrey E. Antipov}
\affiliation{Microsoft Quantum, Microsoft Station Q, University of California, Santa Barbara, California 93106, USA}

\author{Mark van Schilfgaarde}
\affiliation{Department of Physics, King's College London, Strand Campus, London WC2R 2LS, UK}
\affiliation{National Renewable Energy Laboratory, Golden, Colorado 80401, USA}

\author{Roman M. Lutchyn}
\affiliation{Microsoft Quantum, Microsoft Station Q, University of California, Santa Barbara, California 93106, USA}

\begin{abstract}
We investigate transport properties of ballistic magnetic Josephson junctions and establish that suppression of supercurrent is an intrinsic property of the junctions, even in absence of disorder. By studying the role of ferromagnet thickness, magnetization, and crystal orientation we show how the supercurrent decays exponentially with thickness and identify two mechanisms responsible for the effect: (i) large exchange splitting may gap out minority or majority carriers leading to the suppression of Andreev reflection in the junction, (ii) loss of synchronization between different modes due to the significant dispersion of the quasiparticle velocity with the transverse momentum. Our results for Nb/Ni/Nb junctions are in good agreement with recent experimental studies. Our approach combines density functional theory and Bogoliubov-de Gennes model and opens a path for material composition optimization in magnetic Josephson junctions and superconducting magnetic spin valves.
\end{abstract}

\keywords{
	Magnetic Josephson junction,
	$\pi$-junction,
	JMRAM,
	\textit{ab initio} calculations,
	scattering theory.
}

\maketitle

\section{Introduction} \label{sec:introduction}

Coherent quantum tunneling of Cooper pairs through a thin barrier is one of the first examples of macroscopic quantum coherent phenomena. Predicted by Josephson more than 50 years ago~\cite{Josephson:1962}, it has important applications in quantum circuits used in metrology, quantum sensing and quantum information processing~\cite{Warburton:2011}. 

Most of the previous studies focused on conventional Josephson junctions (JJs) consisting of two $s$-wave superconductors (S) that are connected by an insulating (I) or a normal (N) region~\cite{Beenakker:1992, Golubov:2004}. The flow of supercurrent through a JJ depends on the superconducting phase difference $\phi$ between two superconductors and, in general, is characterized by the current-phase relationship $J(\phi)$ (CPR). In conventional JJs CPR should be periodic with $2\pi$, $I(\phi) = I(\phi + 2\pi)$ which follows from the BCS theory~\cite{Bardeen:1957}. This result is a manifestation of a $2e$ charge of Cooper pairs, and is used in metrology to measure electron charge. Time-reversal symmetry requires that $I(\phi) = -I(-\phi)$ which imposes a constraint that the supercurrent should be zero for $\phi = \pi n$ where $n$ is integer. In general, CPR can be expanded in Fourier harmonics, $I(\phi) = \sum_n I_n \sin(n\phi)$.

In many cases, however, CPR is well approximated by the first harmonic $I(\phi) \approx \Ic \sin(\phi)$ with $\Ic$ being the maximum supercurrent that can flow through the junction, i.e. the critical current. At a microscopic level, the supercurrent through a short SNS junction is determined by bound states forming in the constriction due to Andreev reflection at the NS interfaces. In the Andreev reflection process an incident electron-like quasiparticle with spin $\uparrow$ gets reflected at the NS interface as a hole-like quasiparticle with spin $\downarrow$ and a Cooper pair is emitted into the condensate. When time-reversal symmetry is not broken, electrons and holes propagate with the same velocity in the normal region. In this case no phase shift accumulates between this pair of quasiparticles along their trajectories in the N region, and the sign of $\Ic$ is fixed. When $\Ic > 0$ we refer to this case as $0$-junction.

\begin{figure}
	\centering
	\begin{subfigure}[t]{0\linewidth} \phantomcaption \label{fig:sfs}\end{subfigure}%
	\begin{subfigure}[t]{0\linewidth} \phantomcaption \label{fig:sfnfs} \end{subfigure}%
	\begin{subfigure}[t]{\linewidth} \includegraphics[width=7.5cm]{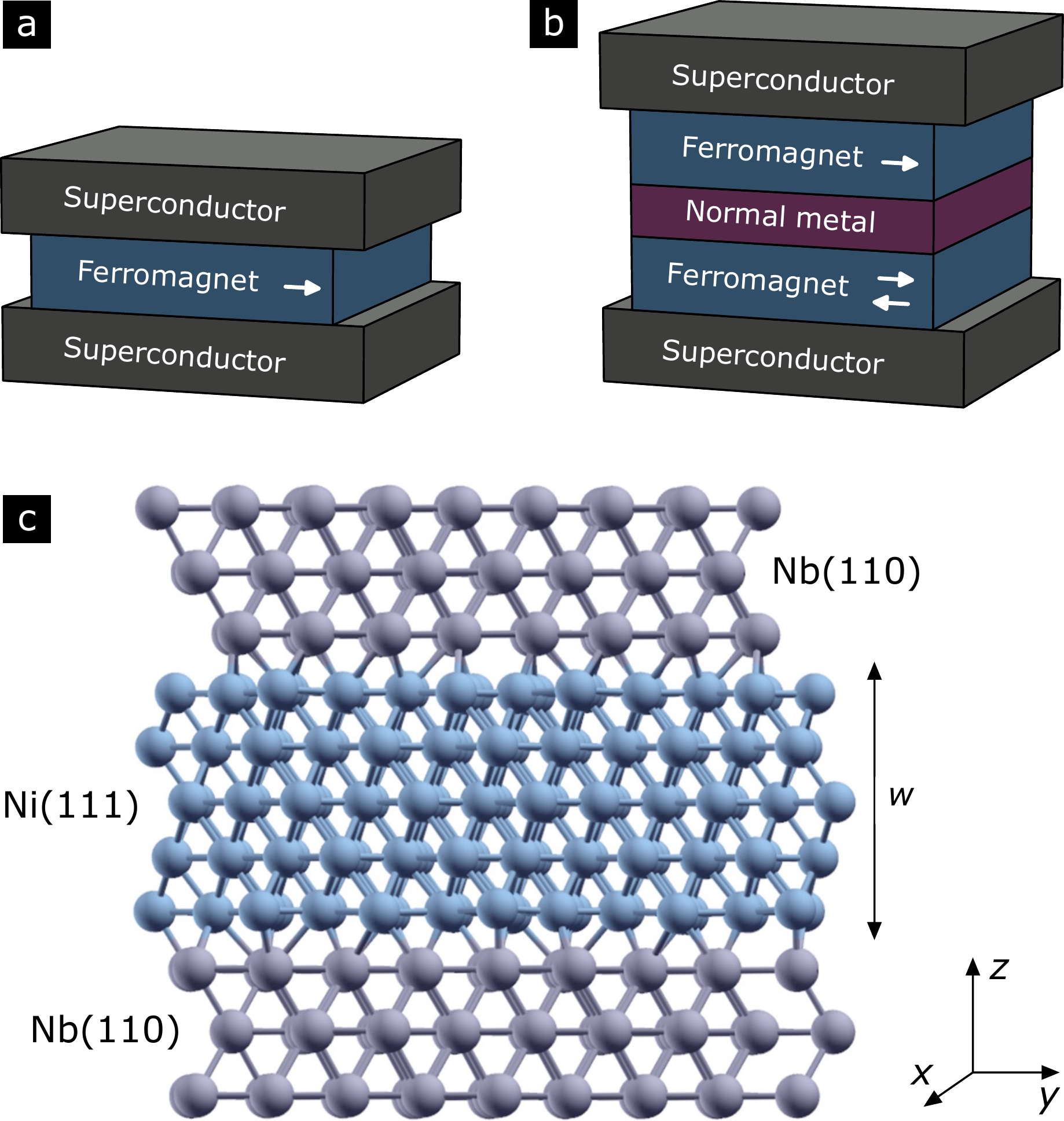} \phantomcaption \label{fig:nbni111nb} \end{subfigure}
	\caption{ \label{fig:stacks}
		\subref{fig:sfs}~Schematic view of the SFS junction.
		\subref{fig:sfnfs}~SFNFS junction. Arrows indicate possible magnetization of
		ferromagnets. The supercurrent through the spin-valve JJ depends on the relative
		magnetization of the ferromagnets, which governs the properties of the Josephson
		magnetic random-access memory (JMRAM)~\cite{Dayton:2018}.
		\subref{fig:nbni111nb}~Ball-and-stick representation of the
		Nb(110)/Ni(111)/Nb(110) junction with 5 layers of Ni. Nb atoms are light grey,
		Ni atoms are blue. The top and bottom atomic planes of Nb(110) are repeated
		periodically in the $z$-direction to create the semi-infinite Nb leads of the
		junction through which the current flows. Periodic boundary conditions are used
		in the $xy$-plane. The corresponding reciprocal space defines two-dimensional
		$\kpar = (k_x, k_y)$ vectors, i.e. the transverse modes, used in the
		calculations.
	}
\end{figure}

In a magnetic Josephson junction (MJJ), exchange splitting breaks time-reversal symmetry and leads to an interesting interplay of superconductivity and magnetism~\cite{Golubov:2004, Buzdin:2005, Bergeret:2005, Blamire:2014, Eschrig:2015}. In superconductor-ferromagnet-superconductor (SFS) junctions [Fig.~\ref{fig:sfs}] the correlated quasi-particles and quasi-holes forming Andreev bound states propagate through the junction under the exchange field of the ferromagnet (F). In many ferromagnets, such as Fe or Ni, the exchange splitting is large (of the order of eV) and significantly perturbs the band structure of a metal and, consequently, significantly modifies Fermi velocities of minority (spin $\downarrow$) and majority (spin $\uparrow$) carriers. Strong time-reversal symmetry breaking leads to the appearance of characteristic superconducting correlations with oscillatory dependence determined by the difference in wave numbers, $\kF^\uparrow - \kF^\downarrow$~\cite{Fulde:1964, Larkin:1965}. This effect opens a possibility for the supercurrent reversal as a function of the thickness of the ferromagnetic region, the so-called Josephson $\pi$-junction~\cite{Buzdin:1982, Ryazanov:2001}. The correlation between the phase shift of the supercurrent and the magnetization provides a possibility for realizing magnetic spin valves, see Fig.~\ref{fig:sfnfs}, which may have promising novel applications for cryogenic superconducting digital technologies~\cite{Bell:2004, Gingrich:2016, Dayton:2018}. Understanding the microscopic physics of MJJs is of great scientific interest as well as technological importance. $0$-$\pi$ transitions in SFS junctions have been extensively studied experimentally since the early 2000's and have been observed in different material systems~\cite{Ryazanov:2001, Ryazanov:2001prb, Kontos:2002, Sellier:2003, Robinson:2006, Khaire:2009, Baek:2017, Baek:2018, Birge:2020, Mishra:2021}. While qualitatively these observations are consistent with the previous phenomenological theories~\cite{McMillan:1968, Wolfram:1968, Kulik:1970, Demers:1971, Griffin:1971, Entin:1977, Blonder:1982, Furusaki:1991, Furusaki:1992, Furusaki:1994, deJong:1995, Tanaka:1997, Zutic:1999, Radovic:2003, Cayssol:2005, Konschelle:2008, Tzortzakakis:2019} the roles of the microscopic band structure arising from the atomic lattice on the supercurrent suppression with junction thickness remain unclear. Our primary goal is to address these essential points. The supercurrent suppression that is exponential in the junction length is often associated with presence of disorder in the ferromagnetic region~\cite{Demler:1997, Buzdin:2005}. However, significant supercurrent suppression can also appear in relatively clean metals (e.g., Ni) whose mean free path is larger than the junction thickness~\cite{Gall:2016, Baek:2017, Baek:2018}.

\begin{figure}
	\centering
	\begin{subfigure}[t]{0\linewidth} \phantomcaption \label{fig:band_sketch} \end{subfigure}%
	\begin{subfigure}[t]{\linewidth} \includegraphics[width=7.5cm]{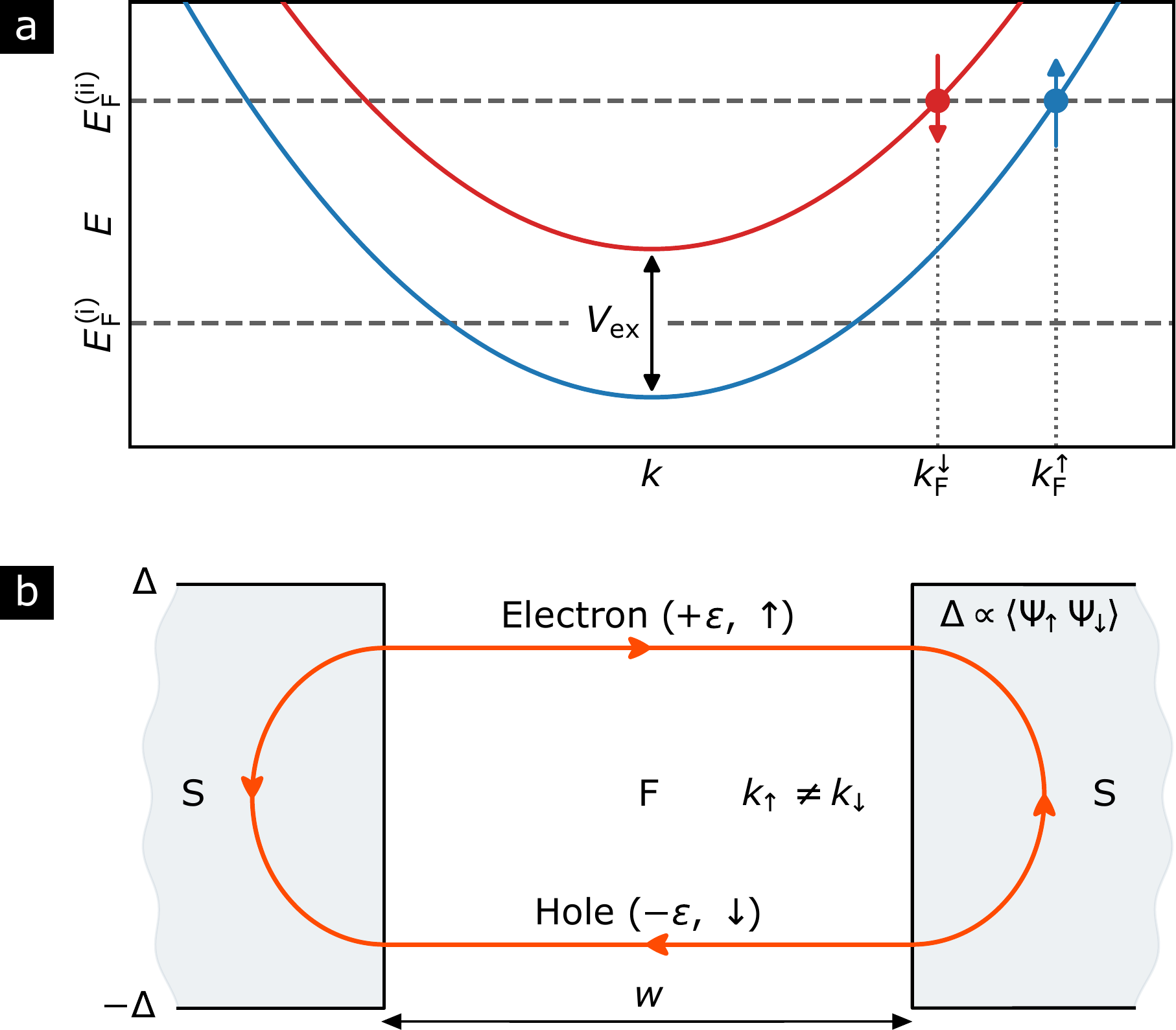} \phantomcaption \label{fig:sfs_sketch} \end{subfigure}
	\caption{ \label{fig:sfs_band_sketch}
		\subref{fig:band_sketch}~Simplified band structure of a ferromagnet, with
		majority and minority bands split by $\Vex$. $\kF^\uparrow$ and $\kF^\downarrow$
		are the Fermi momenta for majority and minority carriers, respectively. Fermi
		level $\EF^\mathrm{(i)}$ corresponds to large $\Vex$, where the minority band is
		pushed above $\EF^\mathrm{(i)}$. $\EF^\mathrm{(ii)}$ corresponds to small
		$\Vex$. Thus, the propagation of minority quasiparticles is characterized by an
		imaginary momentum $\kF^\downarrow$ and is suppressed.
		\subref{fig:sfs_sketch}~Supercurrent in SFS junction is carried by Andreev bound
		states localized in the junction. Solid red line represents quasi-classical
		trajectory corresponding to an Andreev bound state. The spectrum of Andreev
		states depends on the relative superconducting phase difference across the
		junction as well as the phase, $\delta\varphi = |\kF^\uparrow - \kF^\downarrow| w$,
		accumulated due to the difference of Fermi momenta for majority and
		minority carriers. Note that in scenario~(i) the propagation of minority
		carriers is suppressed leading to an overall exponential decay of the
		supercurrent with $w$. This is to be contrasted with the normal transport
		through the junction.
	}
\end{figure}

\begin{figure*}
	\centering
	\begin{subfigure}[t]{0\linewidth} \phantomcaption \label{fig:J1fit} \end{subfigure}%
	\begin{subfigure}[t]{0\linewidth} \phantomcaption \label{fig:G} \end{subfigure}%
	\begin{subfigure}[t]{0\linewidth} \phantomcaption \label{fig:J_phi_4L} \end{subfigure}%
	\begin{subfigure}[t]{0\linewidth} \phantomcaption \label{fig:J_phi_8L} \end{subfigure}%
	\begin{subfigure}[t]{\linewidth} \includegraphics[width=17.5cm]{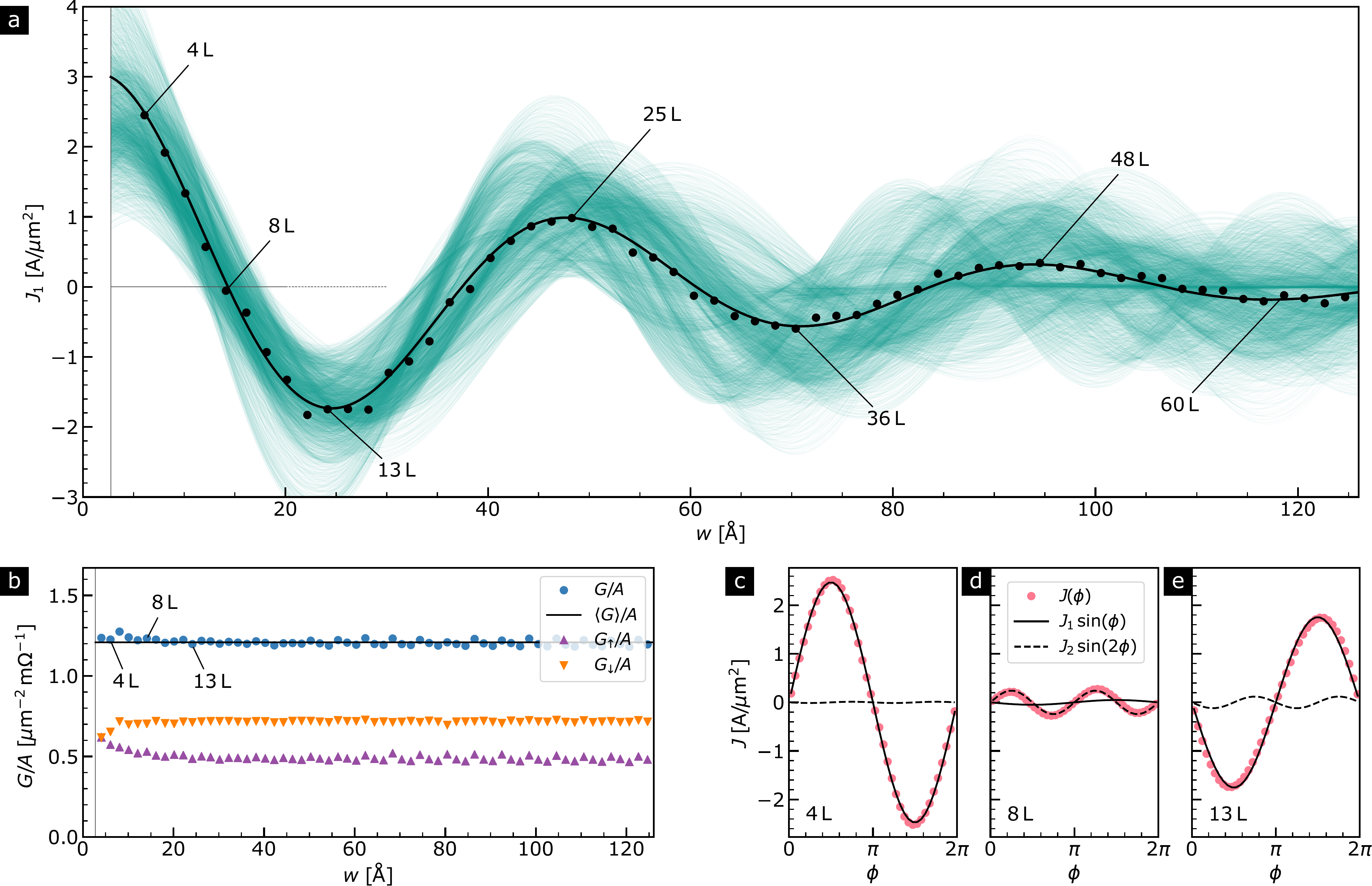} \phantomcaption \label{fig:J_phi_13L} \end{subfigure}
	\caption{ \label{fig:J1_G_w}
		\subref{fig:J1fit}~First Fourier harmonic $J_1$ of the supercurrent density
		[Eq.~\eqref{eq:J}, black circles] and its fit [Eq.~\eqref{eq:J1fit}, solid black
		line] as a function of Ni layer thickness, $w$, calculated for the
		Nb(110)/Ni(111)/Nb(110) junctions shown in Fig.~\ref{fig:nbni111nb}. Green
		semitransparent curves correspond to $j_1^\mathrm{fit}(\kpar)$ for individual
		$\kpar$.
		\subref{fig:G}~Normal state conductance per unit of area as a function of $w$
		for majority ($\Gup$) and minority ($\Gdn$) spins [Eq.~\eqref{eq:G}] as well as
		$\G = \Gup + \Gdn$. $\G$ does not depend on $w$ and is approximated by the
		single value $\langle \G \rangle$.
		\subref{fig:J_phi_4L}--\subref{fig:J_phi_13L}~Supercurrent as a function of
		phase difference $\phi$ for
		\subref{fig:J_phi_4L}~4 atomic layers of Ni (strong $0$-junction regime),
		\subref{fig:J_phi_8L}~8 layers (intermediate regime), and
		\subref{fig:J_phi_13L}~13 layers (strong $\pi$-junction regime). In the $0$- and
		$\pi$-junction regimes, the $J_1$ component dominates. In the intermediate
		regime higher-order terms may prevail.
	}
\end{figure*}

Here we study the suppression of the critical current in the MJJs shown in Fig.~\ref{fig:sfs} and identify two microscopic mechanisms for its suppression, both a consequence of the band structure asymmetry of the majority and minority carriers in the F region. First, there is an asymmetry in the structure of the Fermi surface, see Fig.~\ref{fig:band_sketch}. For certain bands and momenta, the Fermi surface present in one spin channel may be absent in the other. This is typical in ferromagnetic materials like Fe, Co, and Ni because the bandwidth for $d$-electrons is relatively small and is often comparable to the exchange splitting. As a result, the wave number of one of the constituent quasi-particles forming Andreev bound states in MJJ becomes imaginary and the supercurrent becomes suppressed. We label this scenario as mechanism (i). In a second mechanism (ii), we show there is a dephasing of a harmonic signal originating from different Fourier components of the supercurrent due the Fermi velocity dispersion, see Fig.~\ref{fig:J1fit}. Both these mechanisms lead to an exponential suppression of the supercurrent, which was previously believed to occur due to the presence of disorder in the magnetic region. We show that band structure effects are important and may be even dominant in many cases.

In order to capture realistic band structure, we develop a microscopic theory for the supercurrent in realistic MJJs. We use a combination of density functional theory (DFT) and Bogoliubov-de Gennes (BdG) model to investigate the $0$-$\pi$ transition in realistic material stacks of Nb/Ni/Nb junctions in the clean limit. This method allows one to predict and explain key properties of MJJs such as the period and decay of the critical current oscillations with the ferromagnet thickness.

The paper is organized as follows. We describe our findings and summarize our main results in Sec.~\ref{sec:executive_summary}. In Sec.~\ref{sec:method} we discuss the numerical method we developed to perform first principles supercurrent calculations. Detailed discussion of the main results is presented in Sec.~\ref{sec:results}. We conclude with Sec.~\ref{sec:conclusion}. Some technical points are relegated to Appendices~\ref{app:mean_free_path}--\ref{app:scattering_matrix_calculation}.

\section{Qualitative discussion and main results} \label{sec:executive_summary}

In this section we describe basic concepts for the supercurrent flow in MJJs and summarize our results. Our main qualitative conclusions are supported by microscopic calculations for Nb/Ni/Nb MJJs. Ni appears to have fairly long mean free path $\lMFP \approx 60$\,{\AA} (see estimations in Appendix~\ref{app:mean_free_path}), which is comparable or larger than the typical thickness of the ferromagnet used in recent experiments~\cite{Baek:2017, Baek:2018}. Therefore, the motion of quasiparticles in Nb/Ni/Nb junction is quasi-ballistic, and our method is applicable to this system. Most of this paper focuses on clean Nb/Ni/Nb junctions.

First it is illuminating to consider a toy model, a one-dimensional SFS junction, and calculate the supercurrent in such a system for different Fermi energies, see Fig.~\ref{fig:band_sketch}. Using the results of Ref.~\onlinecite{Cheng:2012}, one finds that both majority and minority spin bands are both occupied in the limit $\Vex/\EF \ll 1$ [i.e. scenario~(ii) in Fig.~\ref{fig:band_sketch}] the supercurrent does not decay with ferromagnet thickness at zero temperature,
\begin{align}
	I(\phi)
	= \frac{2e\Delta}{\hbar} \begin{cases}
		\cos\delta\varphi\sin\cfrac{\phi}{2}, & 0<\phi<\pi-2\delta\varphi, \\
		-\sin\delta\varphi\cos\cfrac{\phi}{2}, & \pi-2\delta\varphi<\phi<\pi+2\delta\varphi, \\
		-\cos\delta\varphi\sin\cfrac{\phi}{2}, & \pi + 2\delta\varphi<\phi<2\pi.
	\end{cases}
	\label{eq:Ic0a}
\end{align}
Here perfect interface transparency $\mathcal{T}$ is assumed, $\mathcal{T} \approx 1$. The phase offset $\delta\varphi$ originates from the Fermi momentum difference of a quasi-particle and a quasi-hole forming Andreev bound state in the junction, see Fig.~\ref{fig:sfs_sketch}, and is given by $\delta\varphi=|\kF^\uparrow - \kF^\downarrow|w \approx \Vex w / \hbar \vF$. The $0$- and $\pi$-junction regimes can be clearly identified at $\delta\varphi = 0$ and $\delta\varphi = \pi/2$, respectively. At the intermediate values $0 < \delta\varphi < \pi/2$, the CPR is anharmonic which is a generic feature at $0$-$\pi$ transition as shown below. In the low transparency regime, $\mathcal{T} \ll 1$, one would expect qualitatively similar results with the maximal critical current being suppressed $\Ic \sim (e\Delta / \hbar) \mathcal{T}$ but still independent of the ferromagnet thickness, $w$.

In the case of large exchange splitting $\Vex / \EF \gg 1$ [scenario~(i) in Fig.~\ref{fig:band_sketch}] the minority band may become unoccupied. Given that minority carriers are gapped out and their propagation through the junction is suppressed, the supercurrent decays exponentially with $w$~\cite{Cheng:2012},
\begin{equation}
	I(\phi)
	\approx \frac{2e\Delta}{\hbar} \exp(-\kappa w)
	\Bigl[1 - \frac{\EF}{8 \Vex}\sin^2(k w) \Bigr] \sin\phi.
	\label{eq:Ic0b}
\end{equation}
Here $\kappa = \sqrt{2m^*(\Vex \!-\! \EF)} / \hbar$ and $k = \sqrt{2m^*(\Vex \!+\! \EF)} / \hbar$ with $m^*$ being effective electron mass. One may notice the drastic difference between normal-state and superconducting transport in this case --- the former is weakly affected (because majority and minority contributions are additive) whereas the supercurrent is strongly suppressed. In this case, the measurement of normal-state junction resistance does not necessarily predict the magnitude of the supercurrent through the junction.

We now generalize above results for the realistic three-dimensional (3D) geometry and material composition of the MJJ. In the clean limit, the supercurrent $I(\phi)$ in the short-junction limit ($w$ much smaller than the coherence length of the superconductor) is obtained from the spectrum of the Andreev bound states $\varepsilon_\nu(\phi, \kpar)$ localized in the junction~\cite{Beenakker:1992} which now also depends on the parallel momentum $\kpar$. The supercurrent density $J(\phi)$ per junction area $A$ is $J(\phi) = I(\phi) / A$. For the junction with periodic atomic structure in $xy$-plane the supercurrent density at zero temperature is given by
\begin{equation}
	J(\phi)
	= - \frac{e}{\hbar}
	\int\limits_\BZ \! \frac{d\kpar}{(2\pi)^2} \,
	\sum_{\nu > 0}
	\frac{\partial\varepsilon_\nu(\phi, \kpar)}{\partial\phi},
	\label{eq:J}
\end{equation}
where the $\kpar$ integration is performed over the Brillouin zone (BZ) of the corresponding surface supercell of area, $A$, and the sum is carried over positive quasiparticle energies, $\varepsilon_\nu(\phi, \kpar) > 0$. Note that we use spin-resolved $\varepsilon_\nu$ and therefore omit spin prefactor 2 in Eq.~\eqref{eq:J}. The derivative is periodic in $\phi$ and can be represented as a Fourier series
\begin{equation*}
	- \frac{e}{\hbar} \,
	\frac{\partial\varepsilon_\nu(\phi, \kpar)}{\partial\phi}
	= \sum_{n \geqslant 1}
	I_{n\nu}(\kpar) \sin\bigl[n\phi + \delta\varphi_{n\nu}(\kpar)\bigr],
\end{equation*}
so that Eq.~\eqref{eq:J} can be written as
\begin{equation*}
	J(\phi)
	= \int\limits_\BZ \! \frac{d\kpar}{(2\pi)^2} \,
	\! \sum_{\nu > 0} \,
	\sum_{n \geqslant 1}
	I_{n\nu} \sin\bigl[n\phi + \delta\varphi_{n\nu}(\kpar)\bigr].
\end{equation*}
As we will show below, away from $0$-$\pi$ transition the supercurrent is dominated by the first ($n = 1$) harmonic. Therefore, we focus henceforth on the first harmonic contribution and drop $n$ index in the following discussion. Next, one may notice that the supercurrent amplitudes $I_{\nu}(\kpar)$ and phase offsets $\delta\varphi_{\nu}(\kpar)$ depend on the parallel momentum $\kpar$. One may include this dependence and define an effective Fermi energy $\EF(\kpar)$ that counts the energy corresponding to $\kpar$ in each band of the ferromagnet from the bottom of the band. Depending on $\EF(\kpar)$ and $\Vex$, either scenario~(i) or (ii) of Fig.~\ref{fig:band_sketch} may be realized.

Thus, it is important to compare the exchange splitting with the bandwidth of $d$-character states in transition metals to make sure that a perturbation theory in $\Vex$ is justified. Assuming it is the case, one can estimate the phase offset $\delta\varphi_{\nu}(\kpar)$ by expanding in exchange splitting to find
\begin{equation}
	\delta\varphi_{\nu}(\kpar)
	\approx \Vex w / \hbar v_{z\nu}(\kpar).
	\label{eq:vperp}
\end{equation}
In general, the dependence of $v_{z\nu}$ on $\kpar$ is complicated, especially in $spd$-transition metals. The combination of complicated amplitude and phase offset dependence on $\kpar$ leads to a non-trivial supercurrent dependence on the ferromagnet thickness $w$. As shown in Fig.~\ref{fig:J1fit}, the critical current decays with $w$ for the Nb(110)/Ni(111)/Nb(110) junctions. We analyze the details of the decay and perform an exponential fit in Sec.~\ref{sec:results}. At small thicknesses, below 30\,{\AA}, this decay originates from the evanescent modes corresponding to gapped out minority or majority carriers which cannot propagate through the junction, see Table~\ref{tab:vFermi}. This is the mechanism (i) discussed above. At larger thicknesses (i.e. $w \gtrsim 50$\,{\AA}) a loss of synchronization between different modes due to the dispersion of $v_{z\nu}(\kpar)$ becomes important. This second mechanism (ii) has been previously discussed in the literature~\cite{Buzdin:1982, Buzdin:2005} under assumptions of a single spherical Fermi surface and a small uniform exchange splitting $\Vex$ in the magnetic region. Within these assumptions, one finds that critical current should decay algebraically with the thickness $w$ of a magnetic layer~\cite{Buzdin:1982}. However, as we show below most of these assumptions do not apply to realistic SFS junctions involving transition metals. Thus, in order to understand CPR in realistic MJJs, one needs to use accurate \textit{ab initio} methods, which capture the physical effects described above.

\begin{figure}
	\centering
	\begin{subfigure}[t]{0\linewidth} \phantomcaption \label{fig:bands} \end{subfigure}%
	\begin{subfigure}[t]{\linewidth} \includegraphics[width=8.6cm]{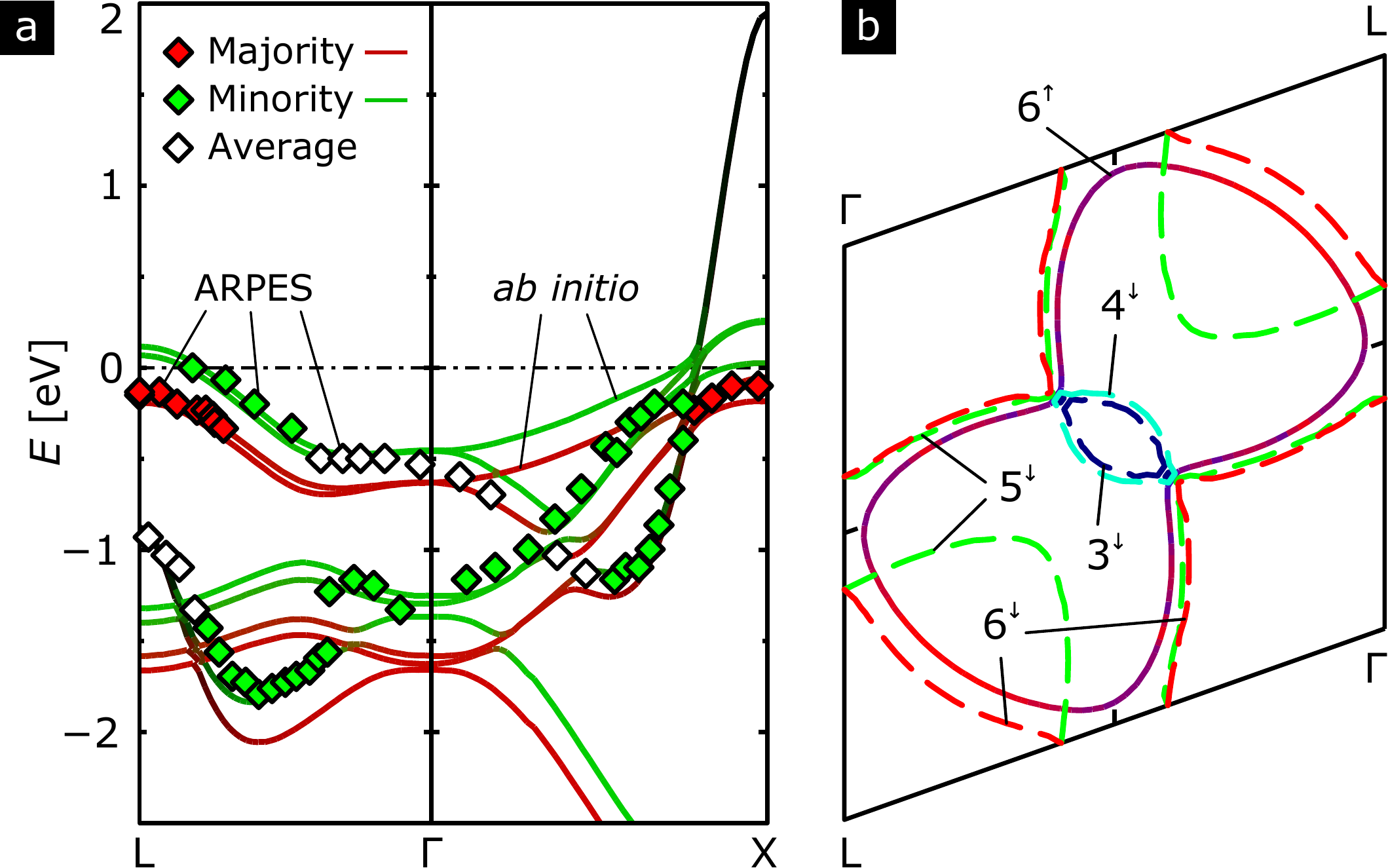} \phantomcaption \label{fig:fsurfaces} \end{subfigure}
	\caption{ \label{fig:bands_fsurfaces}
		\subref{fig:bands}~Electronic band structure of bulk Ni (solid lines) calculated
		from first principles including many-body effects, see
		Ref.~\onlinecite{Sponza:2017}. It is the highest fidelity available and is very
		close to ARPES data (diamonds) in both majority (red) and minority (green) spin
		bands, with exchange splitting $\Vex = 0.3$\,eV.
		\subref{fig:fsurfaces}~Majority (solid line) and minority (dashed line) Fermi
		surfaces of bulk Ni in the $\kpar$ plane (with $k_z = 0$) corresponding to the
		(111) plane direction used for the stacking of the Nb/Ni/Nb junctions, as
		discussed in the text. Axes correspond to two $\Gamma$-L lines: the Fermi
		surface has a three-fold symmetry in the entire plane. Majority band
		$6^{\uparrow}$ is depicted as solid red-blue line with color interpolating
		between red and blue depending on the Fermi velocity $\hbar^{-1} \partial E/
		\partial k$, which ranges between $3\times10^{5}$\,m/s (red) and
		$6\times10^{5}$\,m/s (blue). Minority bands are shown by dashed lines:
		$3^{\downarrow}$ (blue), $4^{\downarrow}$ (cyan), $5^{\downarrow}$ (green), and
		$6^{\downarrow}$ (red).
	}
\end{figure}

To make a connection between the decay seen in Fig.~\ref{fig:J1_G_w} and the mechanisms responsible for it, Fig.~\ref{fig:bands} presents the energy band structure of bulk Ni. It is probably the highest fidelity band structure available: it very closely reproduces ARPES data in both majority and minority spin bands, with exchange splitting $\Vex = 0.3$\,eV~\cite{Sponza:2017}, and should be an excellent predictor of the real Fermi surface and velocities in Ni.

\setlength\extrarowheight{0.04em}
\setlength{\tabcolsep}{0.27em}
\begin{table*}
	\begin{tabular}{c@{\hspace{1.8em}}cccccccc@{\hspace{2.6em}}ccccc}
	\toprule
		& \multicolumn{8}{c@{\hspace{2em}}}{Majority ($\uparrow$)}  &  \multicolumn{5}{c}{Minority ($\downarrow$)}  \\
		\rule{0pt}{1.6em}
		&  $\vF^\mathrm{min}$  &  $\vF^\mathrm{max}$  &  $\langle \vF \rangle$  &  $\sqrt{\langle \vF^2 \rangle}$  & $\rho(\EF)$ &  $\Delta{E}$  &  $m^*\!/m_\mathrm{e}$  & $\kappa$  &  $\vF^\mathrm{min}$  &  $\vF^\mathrm{max}$  &  $\langle \vF \rangle$  &  $\sqrt{\langle \vF^2 \rangle}$ & $\rho(\EF)$  \\
		\rule{0em}{1.1em}
		Band &  [$10^{5}\frac{\mathrm{m}}{\mathrm{s}}$]  &  [$10^{5}\frac{\mathrm{m}}{\mathrm{s}}$]  &  [$10^{5}\frac{\mathrm{m}}{\mathrm{s}}$]  &  [$10^{5}\frac{\mathrm{m}}{\mathrm{s}}$]  & [eV$^{-1}$\AA$^{-3}$] &  [eV]  & & [\AA]  &  [$10^{5}\frac{\mathrm{m}}{\mathrm{s}}$]  &  [$10^{5}\frac{\mathrm{m}}{\mathrm{s}}$]  &  [$10^{5}\frac{\mathrm{m}}{\mathrm{s}}$]  &  [$10^{5}\frac{\mathrm{m}}{\mathrm{s}}$] & [eV$^{-1}$\AA$^{-3}$] \\
	\midrule
		2 &     &     &     &     &       & $-1.59$ & 0.68 & 12 &     &     &     &     & \\
		3 &     &     &     &     &       & $-0.22$ & 3.90 & 13 & 2.7 & 3.6 & 3.3 & 3.6 & 0.004 \\
		4 &     &     &     &     &       & $-0.11$ & 0.61 & 47 & 0.8 & 1.8 & 1.1 & 1.2 & 0.015 \\
		5 &     &     &     &     &       & $-0.10$ & 2.95 & 22 & 0.6 & 3.0 & 1.5 & 1.8 & 0.173 \\
		6 & 3.5 & 6.0 & 4.6 & 5.2 & 0.029 &         &      &    & 0.6 & 3.0 & 2.3 & 2.6 & 0.045 \\
	\bottomrule
	\end{tabular}
	\caption{
		Minimum, maximum, average, and root mean square of the Fermi velocities $\vF$
		for the majority and minority Fermi surfaces in bulk Ni. $\rho(\EF)$ is the
		density of states at the Fermi level, $\Delta{E}$ is the energy of the valence
		band maximum relative to $\EF$ for the majority bands not crossing $\EF$, $m^*$
		($m_\mathrm{e}$) is the effective (bare) electron mass, and $\kappa$ estimates
		the decay exponent of the evanescent mode for a given band at $\EF$.
	}
	\label{tab:vFermi}
\end{table*}

Here we use this potential to analyze the bulk Ni Fermi surface [Fig.~\ref{fig:fsurfaces}] and Fermi velocities (Table~\ref{tab:vFermi}). Counting from the bottom $s$-band, bands of Ni $d$ character are bands 2 to 6. These bands are nearly full: only majority band 6$^\uparrow$ and minority bands $3^\downarrow$--$6^\downarrow$ cross the Fermi level $\EF$. Only band 6 has both majority and minority carriers at the Fermi surface. Bands 6$^{\uparrow}$ and 6$^{\downarrow}$ have roughly the same shape and the energy splitting is approximately constant and equal to $\Vex$ (see the right panel of Fig.~4 in Ref.~\onlinecite{Sponza:2017}). Beyond this, however, the correspondence between the Ni band structure and a simple parabolic band structure deviate substantially, in two ways that critically affect the analysis. First, the Fermi velocity, $\vF={\hbar}^{-1} (\partial E / \partial k)|_{k=\kF}$, is not fixed even for a single pocket: it varies in band $6^\uparrow$ by a factor of 2 [see Fig.~\ref{fig:bands_fsurfaces} and Table~\ref{tab:vFermi}]. Accordingly the splitting $k^\uparrow_\F - k^\downarrow_\F$ between bands $6^\uparrow$ and $6^\downarrow$ varies by factor of two as expected from the twofold variation in $\vF$. Second, bands $3^\downarrow$--$5^\downarrow$ have no majority counterpart at $\EF$, indicating that the wave number of bands $3^\uparrow$--$5^\uparrow$ is complex. This is the origin for the exponential decay in scenario~(i) in Fig.~\ref{fig:band_sketch} as noted above: a large portion of Andreev levels are carried by Cooper pairs made of single-particle wave functions with one or both of $k_\F^{\uparrow}$ and $k_\F^{\downarrow}$ having an imaginary component. The magnitude of $\mathrm{Im}\,k$ depends on the particular mode and $\kpar$ leading to a distribution of decay exponents. The slowest decay in each of these evanescent modes can be estimated from the distance $\Delta{E}$ of the closest approach to $\EF$ and the effective mass $m^{*}/m_e$, using $\hbar^2k_\mathrm{min}^2/2m^*=\Delta{E}$ and decay $\kappa = 2\pi/\mathrm{Im}(k_\mathrm{min})$, see Table~\ref{tab:vFermi}. ($m^{*}/m_e$ is found to be highly anisotropic, so only the effective transport mass $m^{*} = 3\,[1/m_1^{*} + 1/m_2^{*} + 1/m_3^{*}]^{-1}$ is shown.) $\kappa$ is only a rough measure of the evanescent mode decay for a given band. We discuss in detail the distribution of decay exponents and phase offsets in Sec.~\ref{sec:results}.

Let us now focus on the mechanism (ii) for the supercurrent decay, i.e. loss of synchronization between different transverse modes. This mechanism is well-known in diffusive systems where quasiparticle trajectory is random and thus the phase offset $\delta\varphi$ accumulated along such a trajectory also gets randomized. Thus, upon averaging Eq.~\eqref{eq:J} over different disorder realizations, one ends up with exponentially decaying critical current~\cite{Buzdin:2005}. Previously, such a suppression of the supercurrent with junction thickness, $w$, was often associated with impurity scattering in the ferromagnet. Here we demonstrate that this dephasing mechanism can also appear in clean systems (where quasiparticle trajectory is well defined) due to the dispersion of the velocity $v_{z\nu}$ with an in-plane momentum $\kpar$. Specifically, we find that, in the Nb/Ni/Nb junctions, this mechanism becomes relevant for junctions thicker than $50$\,{\AA}, see Fig.~\ref{fig:J1_G_w} and Sec.~\ref{sec:results}. In SFS junctions the combination of disorder in the ferromagnet, interface scattering as well band-structure-induced dephasing ultimately determines the magnitude of the supercurrent. However, we believe that band structure effects provide an upper bound on the magnitude of the critical current as a function of $w$.

In Sec.~\ref{sec:results}, we present numerical results which support the qualitative discussion presented above.

\section{Method} \label{sec:method}

We develop a numerical method to perform realistic simulations of MJJs using a combination of first-principles DFT and BdG calculations. The former is used to obtain the normal-state properties  (e.g., band structure, Fermi velocities, magnetization) and to calculate the normal scattering matrices through the inhomogeneous 3D realistic junctions. As a next step we take superconductivity into account and calculate supercurrent through the stack assuming the short junction limit.

\subsection{Normal transport: \textit{ab initio} description} \label{sec:normal_transport}

To calculate the normal scattering matrix we use the \href{https://www.questaal.org}{\texttt{Questaal} package} for electronic structure calculations based on the linear muffin-tin orbital (LMTO) method~\cite{Pashov:2020}. It calculates the full non-linear, i.e. non equilibrium, transport properties of an infinite system describing a central (C) region cladded by two semi-infinite left (L) and right (R) leads~\cite{Faleev:2005, Meir:1992}, as represented below:
\begin{equation*}
	\overbrace{\ldots \, |\,\mathcal{L}\,|\,\mathcal{L}\,}^{\textstyle \mathrm{L}} \,|\,
	\overbrace{\,\mathrm{PL}_0 \,|\, \mathrm{PL}_1 \,|\, \ldots \,|\, \mathrm{PL}_{L-1}\,}^{\textstyle \mathrm{C}} \,|\,
	\overbrace{\,\mathcal{R}\,|\,\mathcal{R}\,|\, \ldots}^{\textstyle \mathrm{R}}
\end{equation*}
The LCR system [Fig.~\ref{fig:nbni111nb}] can be partitioned into an infinite stack of principal layers (PLs) which interact only with their nearest-neighbors. This is possible because the screened LMTO structure constants are short-ranged~\cite{Andersen:1984}. In the present case the C region consists of the ferromagnet, plus two layers of Nb at the LC and CR interfaces respectively. This is the range over which the perturbation from C significantly modifies the potential in the L or R region. To construct the Nb/Ni/Nb stack, coincident site lattices for Ni and Nb must be found (details of how this was accomplished are given in Appendix~\ref{app:Ni111}). Planes of coincident site lattices are stacked to form the Nb/Ni/Nb structures. Figure~\ref{fig:supercells} shows the Nb and Ni planes we used, which are denoted here as `surface supercells.'

The electronic current flows along the $z$ direction, perpendicular to the PLs lying in the $xy$-plane (transverse direction), see Fig.~\ref{fig:nbni111nb}. Periodic boundary conditions are used within each PL. The corresponding reciprocal space defines the two-dimensional (2D) $\kpar$ vectors, i.e. the transverse modes, used in the calculations. The $\kpar$ mesh is discretized and integrals over $\kpar$ are performed numerically.

The electronic structure of the C region can be separated from L and R regions through self-energies, $\Sigma_\L$ and $\Sigma_\R$, that modify the Hamiltonian of C region. They are most easily calculated if the potential of each PL in the L or R region are identical all through the bulk region. This is the reason for adding a few Nb layers folded into the C region. Thus the periodically repeating unit cells in the L and R regions can be safely assumed to have the potential of the bulk crystal. To construct the self-energies, the potentials of the PL in an infinite stack are needed. These potentials are functions only of the PL in their own region, and may be calculated in several ways. $\Sigma_\L$ and $\Sigma_\R$ are obtained from `surface' Green's function (a fictitious system which consists of a semi-infinite stack of PL, each with the same potential). Note that the potential of the C region is calculated self-consistently. This is important, as the local moments of Ni are small at the boundary layers, and build up gradually, see Fig.~\ref{fig:magnetization}.

With the potential in hand, the normal-state transport can be calculated using scattering formalism~\cite{Meir:1992, Fisher:1981}. For this, knowing the retarded Green's function, $\GF$, of the junction is sufficient. However, this is not the case for the Josephson current: the individual eigenfunctions are required. Within a Green's function framework, $\GF$ must be organized by normal modes which correspond to the eigenstates of the L and R leads for a prescribed energy $E$. In the PL representation, the Hamiltonian has been discretized into the linear combinations of the LMTO basis functions, and the normal modes are represented as eigenvectors of these basis functions. The Schr\"odinger equation becomes a difference equation in the PL basis functions~\cite{Chen:1989}. Eigenvectors are calculated by solving a quadratic eigenvalue problem~\cite{Chen:1989, Fujimoto:2003}, whose eigenvalues correspond to $\exp(\pm i k_{z,n} a)$, where $a$ is the thickness of the PL. The wave number $k_{z,n}$ of the normal mode $n$ can be complex, but to correspond to a propagating mode $k_{z,n}$ must be real. By solving the equation as a function of the energy $E$, one gets all the eigenvalues and eigenvectors which provide the information needed to construct the self-energies $\Sigma_\L$ and $\Sigma_\R$ (for each $\kpar$ and each spin $\sigma$). Note that, in the mode basis, the imaginary part of the self-energies is proportional to the (band) velocity of the modes, and is diagonal for non-degenerate modes~\cite{Fujimoto:2003, Wimmer:2008}.

The retarded Green's function $\GF$ of the C region (connected to the L and R leads) can be written as a matrix in the normal mode basis,
\begin{multline}
	\GF_{\sigma}(E, \kpar)  =
	\Bigl\{ [ \GF_{\C, \sigma}(E, \kpar) ]^{-1} \\
	- \Sigma_{\L, \sigma}(E, \kpar)
	- \Sigma_{\R, \sigma}(E, \kpar) \Bigr\}^{-1},
	\label{eq:GF}
\end{multline}
where $\GF_\C$ is the Green's function of the isolated C region. In this basis, $\GF$ is decomposed into four blocks,
\begin{equation}
	\GF_\sigma(E, \kpar) = \left[\begin{array}{cc}
		\GF_{\LL, \sigma}(E, \kpar) & \GF_{\LR, \sigma}(E, \kpar) \\
		\GF_{\RL, \sigma}(E, \kpar) & \GF_{\RR, \sigma}(E, \kpar)
	\end{array} \right] \!,
	\label{eq:GF_blocks}
\end{equation}
upon projecting onto the propagating modes of the L and R leads. These four quantities and the mode velocities completely determine the normal state transport properties of the junctions.

The transmission matrices are defined by the off-diagonal parts of Eq.~\eqref{eq:GF_blocks}. More specifically, the transmission coefficients $[t_{\LR, \sigma}]_{nm}$, connecting L and R regions, are given by~\cite{Fisher:1981}
\begin{multline}
	[t_{\LR, \sigma}]_{nm}(E, \kpar)
	= i \sqrt{|[v_{\L, \sigma}]_n(E, \kpar)|} \\
	\times [\GF_{\LR, \sigma}]_{nm}(E, \kpar) \,
	\sqrt{|[v_{\R, \sigma}]_m(E, \kpar)|},
	\label{eq:tLR}
\end{multline}
where $[v_{\L, \sigma}]_n$ and $[v_{\R, \sigma}]_m$ are the velocity matrix elements for propagating modes $n$ and $m$ in the L and R leads, respectively. The transmission matrix $t_\RL$ can be obtained from Eq.~\eqref{eq:tLR} by replacing $\mathrm{L} \leftrightarrow \mathrm{R}$. The reflection coefficients are given by the diagonal blocks of Eq.~\eqref{eq:GF_blocks}. For instance, on the L side
\begin{multline}
	[r_{\LL, \sigma}]_{nn'}(E, \kpar)
	= i \sqrt{|[v_{\L, \sigma}]_n(E, \kpar)|} \\
	\times [\GF_{\LL, \sigma}]_{nn'}(E, \kpar) \,
	\sqrt{|[v_{\L, \sigma}]_{n'}(E, \kpar)|}
	- \delta_{nn'}.
	\label{eq:rLL}
\end{multline}
The reflection matrix $r_\RR$ on the R side is obtained from Eq.~\eqref{eq:rLL} by replacing $\mathrm{L} \leftrightarrow \mathrm{R}$.

We define the normal scattering matrix as
\begin{equation}
	S =
	\left[\begin{array}{cc|cc}
		r_{\LL,\uparrow} & 0 & t_{\LR,\uparrow} & 0 \\
		0 & r_{\LL,\downarrow} & 0 & t_{\LR,\downarrow} \\ \hline
		t_{\RL,\uparrow} & 0 & r_{\RR,\uparrow} & 0 \\
		0 & t_{\RL,\downarrow} & 0 & r_{\RR,\downarrow}
	\end{array} \right] \!,
	\label{eq:S}
\end{equation}
where we omit $E$- and $\kpar$-dependence for brevity.

The linear-response normal conductance, $\Gsigma$, per spin is given by
\begin{equation}
	\frac{\Gsigma}{A}
	= \frac{e^2}{h}
	\int\limits_\BZ \! \frac{d\kpar}{(2\pi)^2} \,
	\sum_{n, m}
	\bigl| [t_{\LR, \sigma}]_{nm}(\EF, \kpar) \bigr|^2.
	\label{eq:G}
\end{equation}
It is calculated at the Fermi energy, $E = \EF$. The total conductance is given by $\G = \Gup + \Gdn$. Figure~\ref{fig:G} shows that $\G$, $\Gup$, and $\Gdn$ are almost independent of the junction thickness, $w$.

\subsection{Superconducting transport: scattering matrix approach} \label{sec:superconducting_transport}

Equation~\eqref{eq:S} is the normal-state scattering matrix for metal-ferromagnet-metal (NFN) structure taking into account reflection at both NF interfaces. To account for superconductivity, we introduce a step-like superconducting pairing potential $\Delta = 3.1$\,meV and use the Andreev approximation to account for electron-hole scattering processes~\cite{Beenakker:1992, Eschrig:2015}. This approach combines the details of the atomic structure of Nb/Ni/Nb and superconductivity within the mean-field approximation. 

The direct contact between S and F layers, as shown in Fig.~\ref{fig:sfs}, would lead to an interaction between them. The back-action of the ferromagnet on the superconductor (i.e. the inverse proximity effect) results in a spatial dependence of the pairing potential near the SF interfaces~\cite{Halterman:2001, Halterman:2002, Csire:2018}. In the case of a clean SFS junction model, the self-consistent BdG calculations have been discussed in Refs.~\onlinecite{Halterman:2004, Barsic:2007, Halterman:2015, Halterman:2016, Alidoust:2020}. In typical experimental systems the superconductor is disordered, so disorder effect on pairing potential needs to be considered, see, e.g., Ref.~\onlinecite{Yagovtsev:2021}. Furthermore, in recent experiments S and F layers are separated by an intermediate spacer layer, which significantly reduces this inverse proximity effect. Thus, to understand inverse proximity effect in realistic SFS devices one would need to include both of the abovementioned ingredients in the model as well as to take into account the inhomogeneous magnetization in the ferromagnet (discussed in the next section) which is outside the scope of this paper. For the sake of clarity, we focus here on a realistic band structure in the ferromagnet and its effect on the supercurrent in the SFS structures. Henceforth, we also neglect the orbital effects of the fringe magnetic field, created by the ferromagnet. 

Since $\Delta \ll \EF$, spin-resolved Andreev reflection at SN$_\L$ and N$_\R$S interfaces is described by
\begin{equation}
	r_\A(\phi) =
	\left[\begin{array}{cc|cc}
		0 & \mathbf{1} \, e^{i\phi/2} & 0 & 0 \\
		\mathbf{1} \, e^{i\phi/2} & 0 & 0 & 0 \\ \hline
		0 & 0 & 0 & \mathbf{1} \, e^{-i\phi/2} \\
		0 & 0 & \mathbf{1} \, e^{-i\phi/2} & 0
	\end{array} \right] \!,
	\label{eq:rA}
\end{equation}
where $\phi$ is the phase difference between left and right superconducting leads and $\mathbf{1}$ is the identity matrix. In the short junction limit, the main contribution to the supercurrent comes from Andreev bound states localized in the junction having energy $\varepsilon_\nu$. The energy spectrum of Andreev states can be obtained using the following equation~\cite{Beenakker:1991},
\begin{equation}
	\alpha(\varepsilon) \!
	\left[\! \begin{array}{cc}
		0 & \!\! r_\A^*(\phi) \\
		r_\A(\phi) \!\! & 0
	\end{array} \!\right] \!\!
	\left[\! \begin{array}{cc}
		S(\EF \!+\! \varepsilon, \kpar) \!\!\! & 0 \\
		0 & \!\!\! S^*(\EF \!-\! \varepsilon, \kpar)
	\end{array} \!\right] \!
	\Psi^\mathrm{in}
	\! = \! \Psi^\mathrm{in},
	\label{eq:qq}
\end{equation}
where $\alpha(\varepsilon) = \sqrt{1-\varepsilon^2/\Delta^2} + i\varepsilon/\Delta$. The vector
$\Psi^\mathrm{in} = [
	\psi_\mathrm{e\uparrow}^{\L\rightarrow},$
	$\psi_\mathrm{e\downarrow}^{\L\rightarrow},$
	$\psi_\mathrm{e\uparrow}^{\R\leftarrow},$
	$\psi_\mathrm{e\downarrow}^{\R\leftarrow},$
	$\psi_\mathrm{h\uparrow}^{\L\rightarrow},$
	$\psi_\mathrm{h\downarrow}^{\L\rightarrow},$
	$\psi_\mathrm{h\uparrow}^{\R\leftarrow},$
	$\psi_\mathrm{h\downarrow}^{\R\leftarrow}
]^\mathrm{T}$ 
corresponds to the electron- and hole-like (e/h) waves in N$_\L$ and N$_\R$ regions incident on the F region from the left ($\rightarrow$) and from the right ($\leftarrow$).

Simulations show that $S$ is weakly-dependent on $E$ in the range $[\EF - \Delta, \EF + \Delta]$. Therefore, we expand $S(E, \kpar)$ in $E - \EF$ and keep only the leading term, i.e. $S(E, \kpar) \approx S(\EF, \kpar) = S(\kpar)$. Using this approximation, one can simplify the quantization condition~\eqref{eq:qq} and reduce it to the matrix eigenvalue problem (see details in Ref.~\onlinecite{vanHeck:2014}). This approach allows one to reliably calculate the Andreev bound states spectrum, $\varepsilon_\nu(\phi, \kpar)$. The zero-temperature supercurrent, $J$, through the junction is given by Eq.~\eqref{eq:J}. Figures~\ref{fig:J_phi_4L}--\ref{fig:J_phi_13L} show $J$ as a function of a phase difference, $\phi$. Figure~\ref{fig:J1fit} shows the first Fourier harmonic of the supercurrent as a function of the junction thickness, $w$.

\section{Results} \label{sec:results}

It is illuminating to compare our numerical simulations for the supercurrent with the experimental measurements involving quasi-ballistic MJJs. As previously discussed, we believe that the Nb/Ni/Nb junctions represent a good model system for which experimental data is readily available~\cite{Gingrich:2016, Baek:2017, Baek:2018}. The best-performing stacks consist of Nb(110)/Cu/Ni(111)/Cu/Nb(110). Cu spacer layers seem to be essential to get strong supercurrent, likely because it prevents intermixing of the Ni and Nb. Our model junction simulates this geometry, via supercells in the plane normal to the stack to account for the lattice mismatch (see Fig.~\ref{fig:stacks} and Appendix~\ref{app:Ni111}), though we do not include the Cu layers. We anticipate that Cu spacers will mainly affect transmission matrix elements rather than the dependence of the supercurrent on ferromagnet thickness, which is the main focus of this work. Furthermore, as discussed before, the Cu spacers will suppress the direct interaction between the ferromagnet and the superconductor and reduce inverse proximity effect justifying Andreev approximation for the boundary conditions, see Eq.~\eqref{eq:rA}. Therefore, we consider only the simplified Nb/Ni/Nb stack and vary the number of layers (atomic planes) of Ni. Additionally, we also consider effect of different crystallographic orientation of the Ni planes and investigate Nb/Ni(110)/Nb junctions in Appendix~\ref{app:Ni110}.

\begin{figure}
	\centering
	\begin{subfigure}[t]{0\linewidth} \phantomcaption \label{fig:supercells} \end{subfigure}%
	\begin{subfigure}[t]{\linewidth} \includegraphics[width=8.6cm]{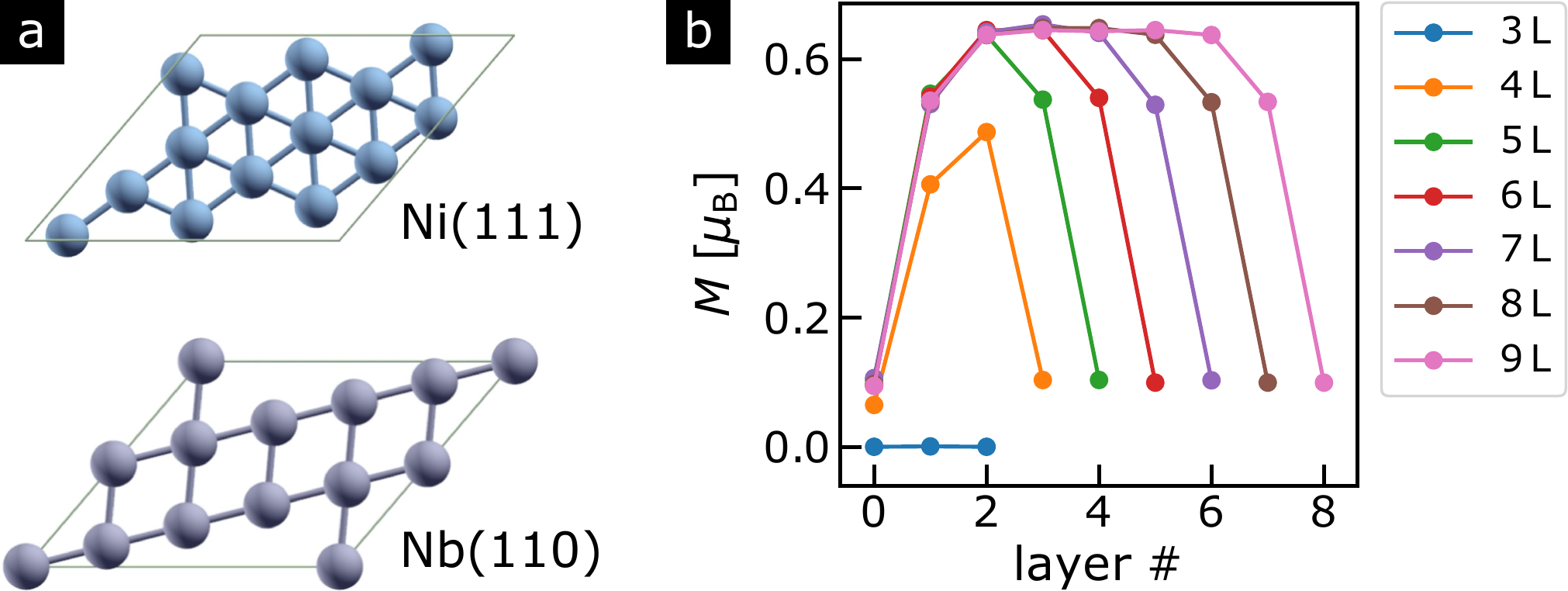} \phantomcaption \label{fig:magnetization} \end{subfigure}
	\caption{ \label{fig:supercells_magnetization}
		\subref{fig:supercells}~Top view of the Nb(110) and Ni(111) surface supercells
		used to build the Nb/Ni interfaces and the Nb(110)/Ni(111)/Nb(110) stacks shown
		in Fig.~\ref{fig:nbni111nb}. The surface supercell is defined from two 2D
		vectors $\mathbf{a}_1 = [10.8, 0]$\,{\AA} and $\mathbf{a}_2 = [6.03,
		7.11]$\,{\AA} with periodic boundary condition in the 2D plane. The
		corresponding reciprocal space defines the 2D $\kpar$ vectors used in the
		calculations. Each atomic plane of Ni (Nb) contains 14 (10) atoms of Ni (Nb).
		\subref{fig:magnetization}~Magnetic moment profile of Nb(110)/Ni(111)/Nb(110)
		junctions for different thickness of Ni (from 3 to 9 layers). The value of the
		moment is an averaged over the moments of the 14 Ni atoms in each atomic plane.
		Note the magnetic dead layer at the Nb/Ni interfaces and that all moments vanish
		for the shortest junction made of 3 layers of Ni.
	}
\end{figure}

To make a Nb/Ni superlattice, the unit cells of the Nb and Ni regions in the plane normal to the interface must be coincident. This is complicated by the severe lattice constant mismatch, and also the incompatibility of the (110) and (111) atomic planes. It is necessary to construct superlattices with Nb(110) and Ni(111) both rotated to the $z$ axis, and with lattice vectors in the plane coincident. A supercell with nearly coincident vectors was found (see Appendix~\ref{app:Ni111} for details). By applying a small shear strain to the Ni, the lattice vectors are made exactly coincident. Figure~\ref{fig:supercells} shows the Nb(110) surface supercell and the Ni(111) surface supercell with equal lattice vectors used to match the Nb/Ni interfaces. Each atomic plane of Ni(111) contains 14 atoms and each atomic plane of Nb(110) consists of 10 atoms. The atomic structure of the Nb(110)/Ni(111)/Nb(110) for 5 layers (atomic planes) of Ni is shown in Fig.~\ref{fig:nbni111nb}.

Next, we performed self-consistent DFT calculations within the local density approximation (LDA) in order to obtain the relaxed structure and corresponding electronic structure. For the smallest structures we performed a constrained optimization. Only the atoms in the planes closest to the Nb/Ni interfaces are allowed to relax to facilitate stacking of arbitrarily large cells. The Nb/Ni interplanar spacing has also been optimized to minimize the total energy, see Appendix~\ref{app:scattering_matrix_calculation} for more details.

Once the structure is determined, one can determine the normal-state thermodynamic and transport properties of the junction, e.g., calculate the magnetization profile and spin-resolved conductance through the junction as a function of Ni thickness. For transport calculations, we use a layer transport technique~\cite{Faleev:2005} which employs the atomic spheres approximation (ASA). Careful checks were made of ASA band structures of elemental Nb and Ni, and also superlattices, to confirm that they are very similar to the full potential LMTO DFT-LDA ones.

We find that the magnetic properties of Ni are sensitive to their local environment, indicative of the itinerant ferromagnetism. As shown in Fig.~\ref{fig:magnetization}, the magnetization profile is non-uniform in the junction with averaged magnetic moments per atom being suppressed near the Nb interface. For thickness larger than 4 layers, one recovers the bulk value of $\sim 0.6 \muB$ in the middle layers, away from the Nb/Ni interfaces. The averaged moment drops down towards the edges and becomes considerably reduced down to $\sim 0.1 \muB$ at the interface with Nb. The strong reduction of magnetism is exemplified for the short junction with 3 layers of Ni where the moments on the Ni atoms have completely vanished. Such a non-uniform magnetic moment dependence in Nb/Ni/Nb junctions affect superconducting properties of the SFS junctions in a non-trivial way. For example, Nb/Ni/Nb junctions thinner than 4 layers of Ni behave as essentially SNS junctions.

It is well known that the LDA tends to overestimate local moments $M$ in itinerant magnets~\cite{Mazin:2004} because spin fluctuations reduce the average moment~\cite{Moriya:1985}, and underestimate $M$ when local moments are very large~\cite{Sponza:2017}. For Ni, LDA yields $M$ in good agreement with the experiment, but this is likely an artifact of an accidental cancellation of errors. Most important for transport is the exchange splitting $\Vex$, which the LDA predicts to be 0.6\,eV, about twice larger than the experimental value of 0.3\,eV~\cite{Himpsel:1979}. It is possible to reproduce both $M$ and $\Vex$ at the same time, but a high-level theory, potentially including spin-orbit coupling, is needed to surmount both kinds of errors inherent the LDA~\cite{Sponza:2017, Bunemann:2008}. The high cost and poor scaling of such a theory is not practical for these junctions, so we elect to stay within the LDA and scale the self-consistently calculated $\Vex$. This was the approach Karlsson and Aryasetiawan used to calculate the spin wave spectra in Ni~\cite{Karlsson:2000}. Scaling of $\Vex$ can be accomplished using different approaches, e.g., by adding some effective magnetic field to simulate the effect of spin fluctuations. Since Ni is a simple case with a nearly linear relation between $M$ and $\Vex$, the band structure hardly depends on the details in which the LDA potential is modified. Here we first perform fully self-consistent calculations. Then, to construct the potential for transport properties, we rescale the spin component of the density by a constant factor, which we denote as $M/M_0$. This enables parametric studies of transport as a function of $\Vex$. $M/M_0 = 0.5$ yields the observed $\Vex = 0.3$\,eV, and we use this scaling unless stated otherwise. 

The conductance per unit of area, $\G/A$, is shown in Fig.~\ref{fig:G}. It is weakly dependent on the thickness, $w$, of the magnetic layer, as expected for a metallic system in the absence of disorder.

\begin{figure}
	\centering
	\includegraphics[width=8.6cm]{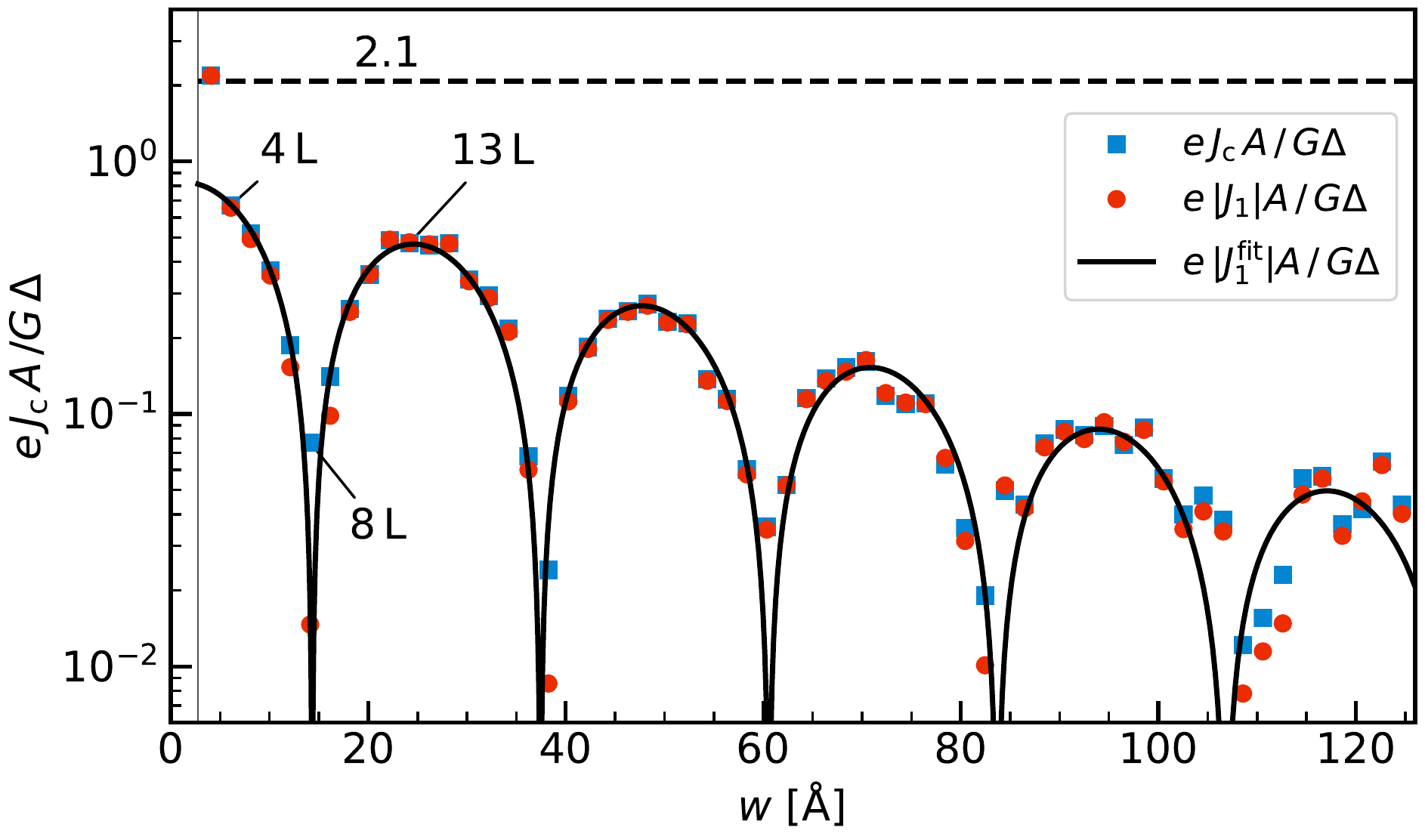}
	\caption{ \label{fig:IcRn_w}
		Comparison of critical current density, $\Jc = \max_\phi |J(\phi)|$ (blue
		squares), absolute value of first Fourier component for supercurrent, $|J_1|$
		[see Eq.~\eqref{eq:J1}, red circles], and its fitting $|J_1^\mathrm{fit}|$
		[Eq.~\eqref{eq:J1fit}, black curve]. All these quantities are `normalized' by
		normal-state conductance, $\G$.
	}
\end{figure}

We now focus on superconducting properties. The dependence of the supercurrent on the phase difference $\phi$ for 5, 8, and 11 Ni layers is shown in Figs.~\ref{fig:J_phi_4L}--\ref{fig:J_phi_13L}. One can present current-phase relation, $J(\phi)$ in a form of a Fourier series,
\begin{equation}
	J(\phi) = \sum\limits_{n \geqslant 1} J_n \sin(n\phi).
	\label{eq:J1}
\end{equation}
In the $0$-junction mode, the first term in the Fourier series dominates with $J_1 > 0$ [see solid black line in Fig.~\ref{fig:J_phi_4L}]. In $\pi$-junction case [Fig.~\ref{fig:J_phi_13L}], the supercurrent is also mostly defined by the first harmonic but with $J_1 < 0$. Close to the $0$-$\pi$ transition $J_1$ dies out, so that the behavior is governed by higher Fourier harmonics~\cite{Stoutimore:2018}, e.g. for 8 Ni layers supercurrent has mostly second harmonic, $J_2$, shown by dashed black line in Fig.~\ref{fig:J_phi_8L}.

Figure~\ref{fig:IcRn_w} shows the critical current density, $\Jc = \max_\phi |J(\phi)|$, normalized by normal-state conductance, $e \Jc A / \G \Delta$, as a function of $w$ (blue squares). Far from the $0$-$\pi$ transitions, the critical current coincides with the absolute value of the first Fourier harmonic, $e |J_1| A / \G \Delta$ (red circles). Since $J_1$ contains a sign of the current and has better numerical stability than $\Jc$, we use this quantity for the analysis. We exclude very thin junctions (3 Ni layers or less) from the analysis since the magnetic properties are suppressed there.%
\footnote{For the first data point in Fig.~\ref{fig:IcRn_w} corresponding to 3 layers of Ni the magnetization is completely suppressed [see Fig.~\ref{fig:magnetization}] and ratio $e \Jc A / \G \Delta \approx 2.2$ is significantly higher than the one for thicker Ni regions with non zero magnetic moments. We compare this value with the result for the short disordered SNS junction. Combination of analytical energy spectrum~\cite{Beenakker:1991} with Dorokhov distribution of channel transmissions~\cite{Dorokhov:1984, Mello:1988} leads to a ratio of 2.1 (horizontal dashed line in Fig.~\ref{fig:IcRn_w}). We attribute this difference to the fact that there is no interfacial disorder in our model.}

\begin{figure*}
	\centering
	\includegraphics[width=17.3cm]{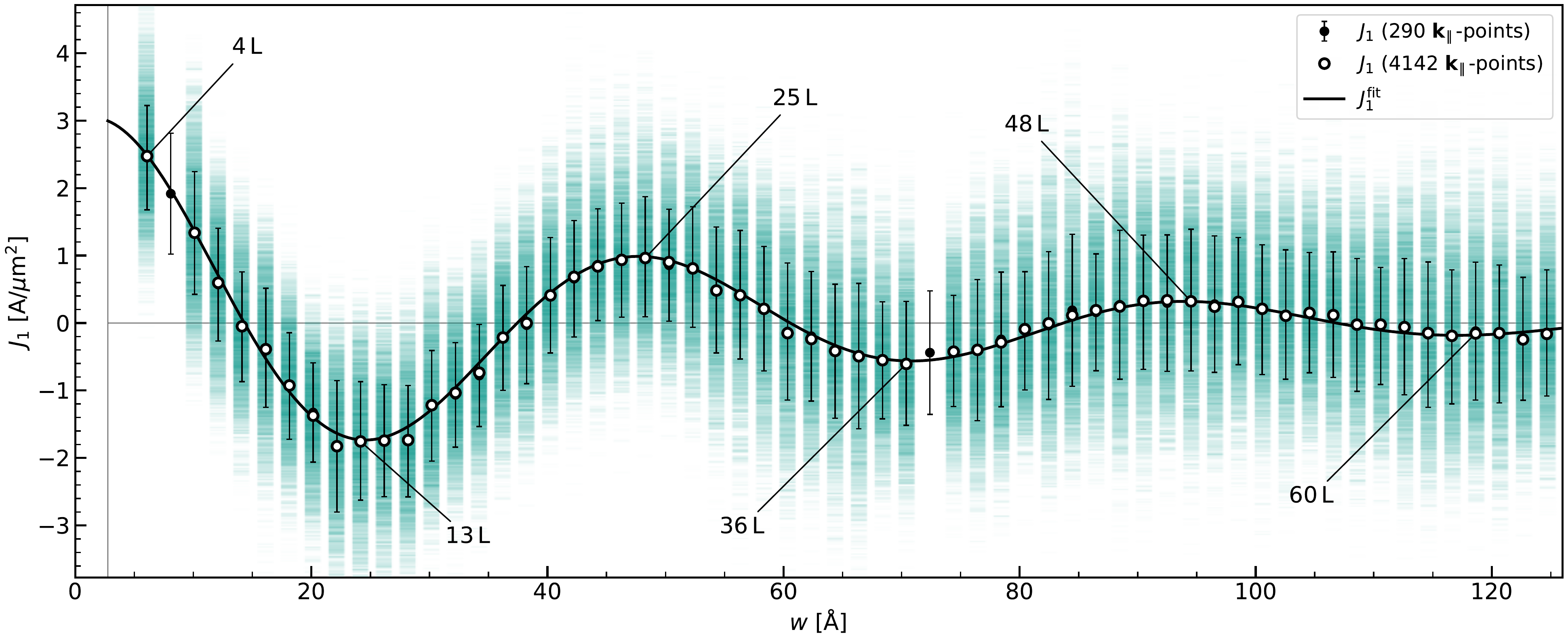}
	\caption{
		First Fourier harmonic, $J_1$, of the supercurrent as a function of junction
		thickness, $w$. Full black circles correspond to $J_1$ calculated with 290
		$\kpar$-points, empty circles (mostly superposed onto the full black circles)
		correspond to 4142 $\kpar$-points. Solid black line is $J_1^\mathrm{fit}(w)$
		[$J_1$ fit given by Eq.~\eqref{eq:J1fit}]. Green semitransparent dashes show
		$J_1^\mathrm{fit}(w)$ contributions [Eq.~\eqref{eq:j1k}] for the individual
		$\kpar$-points. The vertical `errorbars' correspond to the standard deviation of
		$j_1(\kpar)$ with respect to $J_1$. The standard deviation is the same for both
		sets of $\kpar$-point indicating that these results are independent of the
		chosen discretization.
	}
	\label{fig:J1_std_w}
\end{figure*}

The $J_1$ dependence on $w$ can be fit by the following expression,
\begin{equation}
	J_1^\mathrm{fit}(w)
	= \Theta_J \exp(- w / \xi_J)
	\cos\bigl[\pi (w + \delta_J) / \lambda_J \bigr],
	\label{eq:J1fit}
\end{equation}
where $\Theta_J = 3.20\,\mathrm{A}/\mu\mathrm{m}^2$,  $\xi_J = 41.1$\,{\AA}, $\lambda_J = 23.2$\,{\AA}, and $\delta_J = -2.75$\,{\AA} are fitting parameters. We interpret $\xi_J$ as a decay length, $\lambda_J$ as the `half-period' of the oscillation in $J_1$ as a function of $w$, and $\delta_J$ as a measure of the suppressed magnetization in Ni layers near the Ni/Nb boundaries. $J_1^\mathrm{fit}(w)$ accurately fits the discrete points $J_1$ as shown in Figs.~\ref{fig:J1fit}, \ref{fig:IcRn_w}, \ref{fig:J1_std_w}, and \ref{fig:J1_w_M} by black solid line and black circles, accordingly.

\begin{figure*}
	\includegraphics[width=18cm]{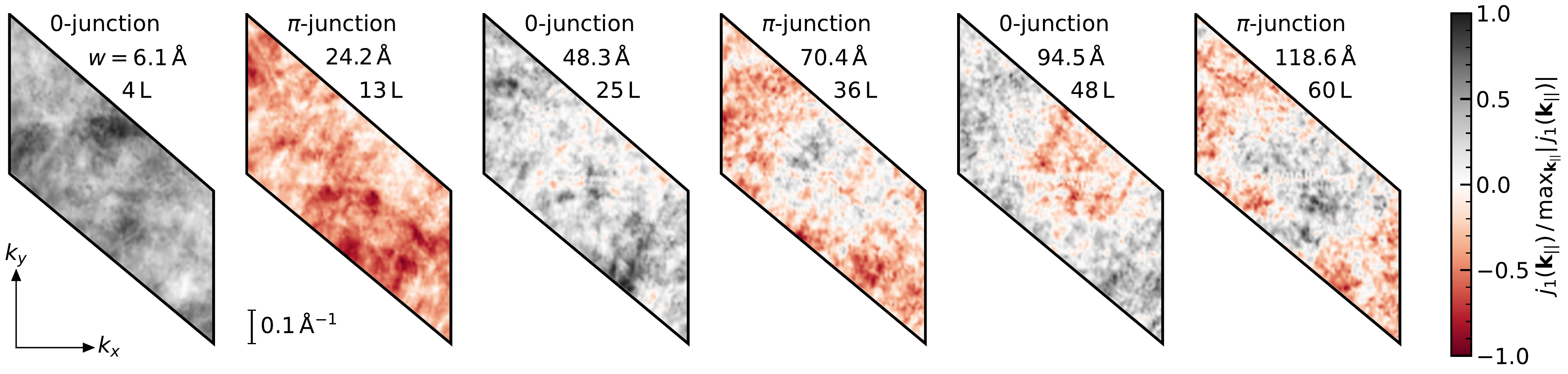}
	\caption{
		Colorplot of $j_1(\kpar)$ with $\kpar = (k_x, k_y)$. Each panel corresponds to
		the local extrema of $J_1^\mathrm{fit}(w)$ shown in Fig.~\ref{fig:J1_std_w}. For
		4 and 13 layers all the $\kpar$ channels contribute with the same sign. For 48
		layers and larger, different $\kpar$ channels lose synchronization and
		contribute to the total supercurrent with different signs.
	}
	\label{fig:k_colormap}
\end{figure*}

In order to gain insight into the evolution of $J_1$ with $w$, let us resolve contributions from different $\kpar$. For this, we rewrite Eq.~\eqref{eq:J} as
\begin{subequations}
\begin{align}
	J(\phi) 
	& = A \int\limits_\BZ \frac{d \kpar}{(2\pi)^2} \,
	j(\phi, \kpar),
	\label{eq:J1_j1k} \\
	j(\phi, \kpar)
	& = - \frac{e}{\hbar} \,
	\frac{1}{A} \,
	\sum_{\nu > 0}
	\frac{\partial\varepsilon_\nu(\phi, \kpar)}{\partial\phi}.
	\label{eq:j1k}
\end{align}
\end{subequations}
Here $A = 76.75$\,\AA$^2$ is the area of the surface supercell shown in Fig.~\ref{fig:supercells}. Similar to Eq.~\eqref{eq:J1}, we denote the first $\phi$-harmonic of $j(\phi, \kpar)$ as $j_1(\kpar)$. The evolution of the first Fourier harmonic of the supercurrent $J_1$ and $j_1(\kpar)$ as a function of $w$ are shown in Figure~\ref{fig:J1_std_w}. Calculations were performed for two different sets of $\kpar$ with 290 discrete $\kpar$-points (full black circles) and 4142 $\kpar$-points (empty black circles). One can see that both sets give the same result for $J_1$, establishing that the $\kpar$ integration is well converged. In Fig.~\ref{fig:J1_std_w}, `errorbars' denote the standard deviation in $j_1(\kpar)$ with respect to the $\kpar$-summed average, $J_1(\phi)$. Individual $j_1(\kpar)$ are shown by semi-transparent horizontal dashes. The important observation is that while $J_1$ decays with $w$, the dispersion in $j_1(\kpar)$ does not change significantly.

Figure~\ref{fig:k_colormap} shows colorplots of $j_1(\kpar)$ corresponding to local extrema of $J_1^\mathrm{fit}(w)$ (4, 13, 25, 36, 48, and 60 layers labeled in Fig.~\ref{fig:J1_std_w}). For small $w$, one can see that most of all the $\kpar$ contributions to $J_1$ have the same sign, i.e. positive in $0$-junction regime and negative in $\pi$-junction regime. In this regime the decay is predominantly due to evanescent modes decaying into the junction. For larger $w$, the dephasing mechanism becomes important since the phase offset spread grows with $w$. One can observe the apparition of contributions of the opposite sign for $w \gtrsim 50$\,{\AA}. This dephasing mechanism is mainly due to the variation of the Fermi velocity with $\kpar$, and becomes more important with increasing $w$.

In order to study the distribution of the phase offsets and decay exponents for different modes, we fit the individual $j_1(\kpar)$ using an expression analogous to Eq.~\eqref{eq:J1fit}. The set of the resulting fitted curves $j_1^\mathrm{fit}(w, \kpar)$ for 4142 $\kpar$-points are shown by the green semitransparent curves in Fig.~\ref{fig:J1fit}. Here to minimize the numerical `noise,' $j_1(w, \kpar)$ curves are smoothed over the 2D $\kpar$ space using Gaussian filter with $\sigma_{\kpar} = 0.01$\,{\AA}$^{-1}$ which is of the order of the Fermi wave vector in Nb. Thus, each data point $j_1(w, \kpar)$ approximately corresponds to a transverse conducting channel. This fitting procedure works reasonably well, e.g., the relationship in Eq.~\eqref{eq:J1_j1k} holds if one replaces $J(\phi)$ by $J_1^\mathrm{fit}(w)$ and $j(\phi, \kpar)$ by $j_1^\mathrm{fit}(w, \kpar)$ for $w \gtrsim 10$\,{\AA}.

\begin{figure}
	\centering
	\begin{subfigure}[t]{0\linewidth} \phantomcaption \label{fig:period_distribution} \end{subfigure}%
	\begin{subfigure}[t]{\linewidth} \includegraphics[width=8.7cm]{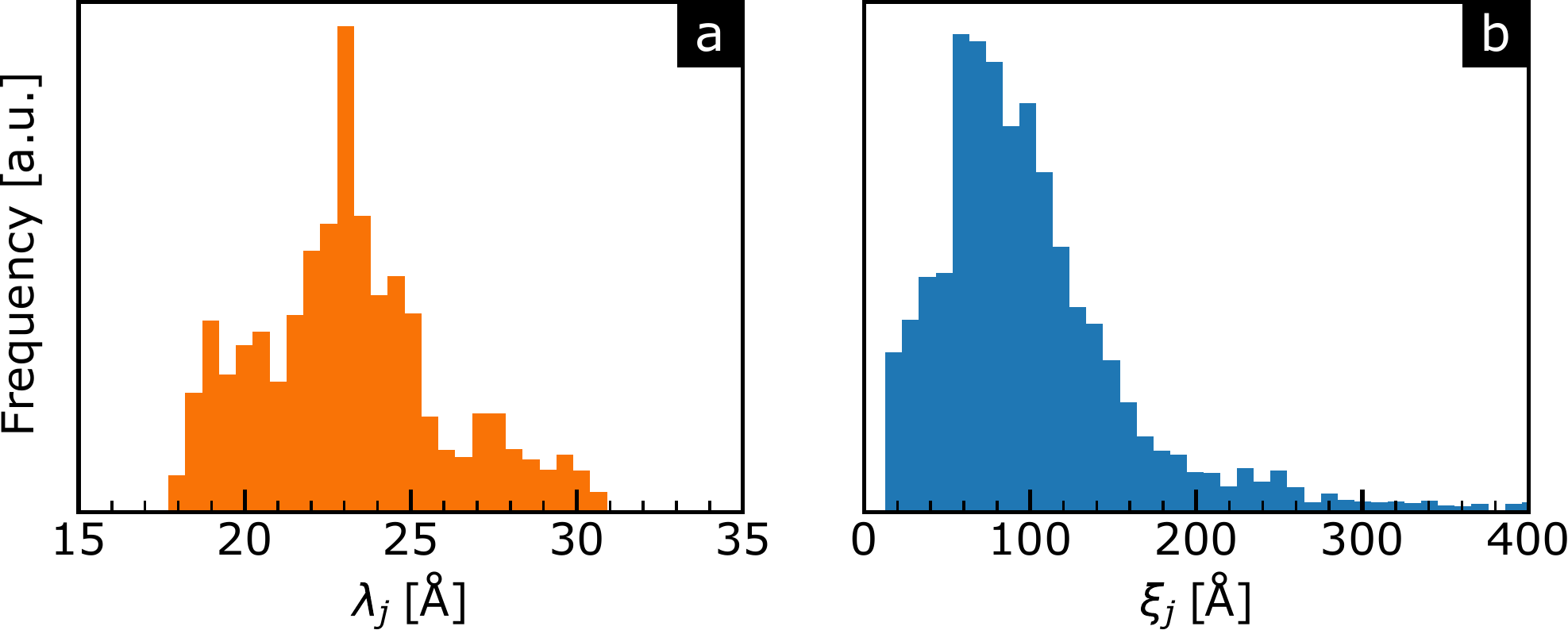} \phantomcaption \label{fig:decay_length_distribution} \end{subfigure}
	\caption{ \label{fig:histograms}
		Analysis of the fitting parameters of $j_1^\mathrm{fit}(w)$ for individual
		$\kpar$, shown in Fig.~\ref{fig:J1fit} by semitransparent green lines.
		Distribution of 
		\subref{fig:period_distribution}~half-periods, $\lambda_j$, and
		\subref{fig:decay_length_distribution}~decay lengths, $\xi_j$.
	}
\end{figure}

\begin{figure}
	\centering
	\begin{subfigure}[t]{0\linewidth} \phantomcaption \label{fig:J1_w_M_orig} \end{subfigure}%
	\begin{subfigure}[t]{\linewidth} \includegraphics[width=8.2cm]{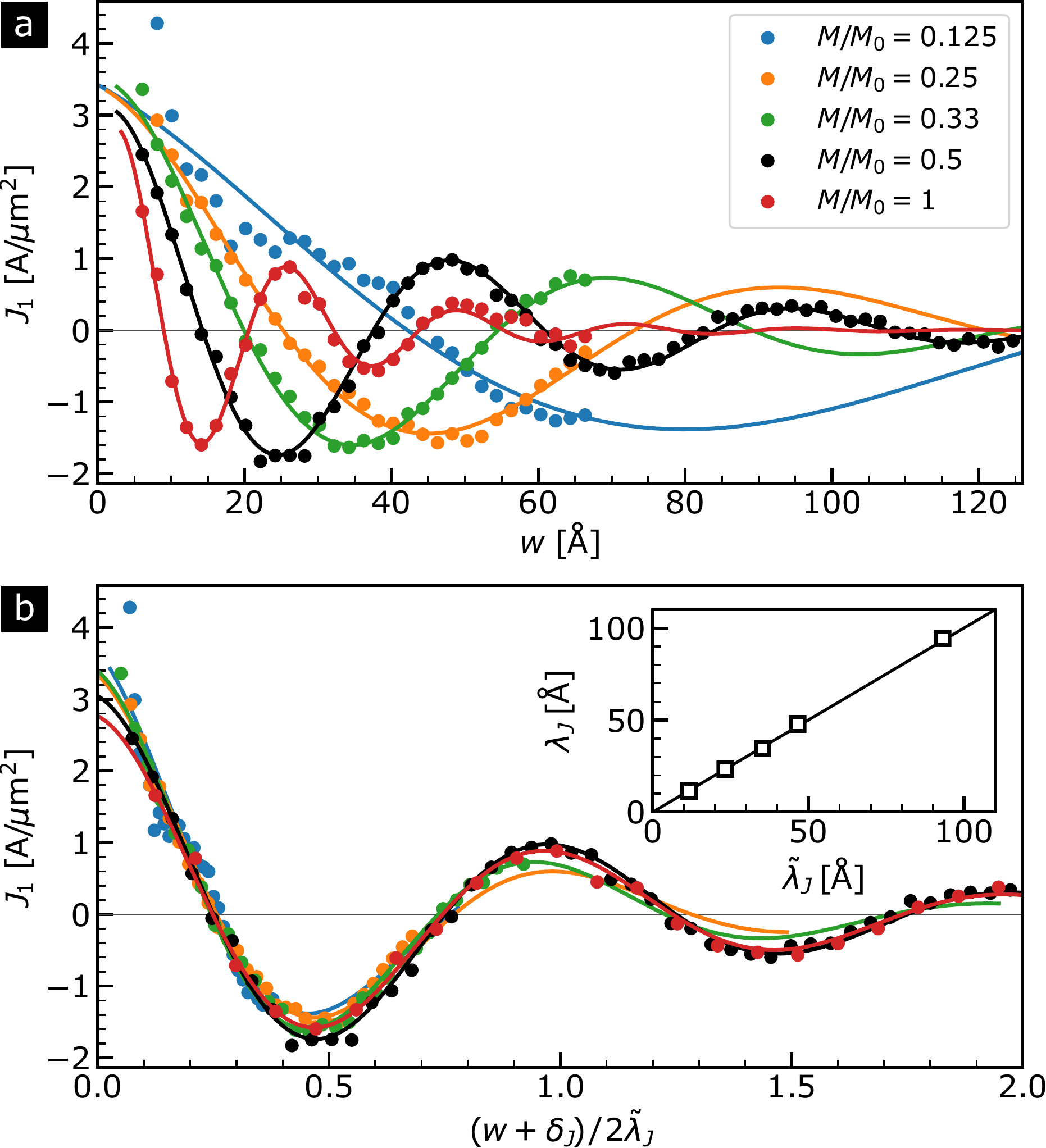} \phantomcaption \label{fig:J1_w_M_resc} \end{subfigure}
	\caption{ \label{fig:J1_w_M}
		$J_1$ for different rescaling of the magnetic moments $M/M_0$ as a function of
		\subref{fig:J1_w_M_orig}~ the thickness, $w$, and
		\subref{fig:J1_w_M_resc}~the rescaled thickness, $(w + \delta_J) /
		2{\tilde\lambda}_J$, where $\delta_J$ is the fitting parameter in
		Eq.~\eqref{eq:J1fit} and ${\tilde\lambda}_J = (M_0/M) \,11.6$\,{\AA}. $J_1$
		values are shown by circles; corresponding fittings $J_1^\mathrm{fit}(w)$ are
		shown by lines of the same color.
		(Inset)~Half-period, $\lambda_J$, fitted using Eq.~\eqref{eq:J1fit} versus
		${\tilde\lambda}_J$.
	}
\end{figure}

The distribution of the fitting parameters for $j_1^\mathrm{fit}(w, \kpar)$ is shown in Fig.~\ref{fig:histograms}. Histograms for decay lengths and half-periods reveal a complicated picture describing different contributions to the supercurrent in real materials. First of all, in Fig.~\ref{fig:period_distribution} one can see the distribution of the half-periods, $\lambda_j$, which is similar to a Gaussian distribution with a mean value $\langle\lambda_j\rangle = 23.2$\,{\AA} and standard deviation $2.8$\,{\AA}. The mean value $\langle\lambda_j\rangle$ is very close to $\lambda_J$ [see text after Eq.~\eqref{eq:J1fit}] while the spread in $\lambda_j$ leads to dephasing and is responsible for the exponential decay of $J_1$ at large $w$. Indeed, it is well-known that the average of an oscillatory function with respect to a random fluctuating phase (described by a Gaussian distribution) results in an exponentially decaying function. 

In addition to the dephasing mechanism, the decay of the supercurrent originates from the evanescent modes. The histogram for decay lengths, $\xi_j$ is shown in Fig.~\ref{fig:decay_length_distribution}. Here small $\xi_j$ corresponds to fast-decaying $j_1^\mathrm{fit}(w, \kpar)$, large $\xi_j$ is responsible for non-decaying modes (i.e. modes with the decay exponents larger than the junction thickness). The right-skewed distribution of the decay exponents has a mean value of $\langle\xi_j\rangle = 108$\,{\AA} which is much larger than $\xi_J$ in Eq.~\eqref{eq:J1fit}. The shoulder at small $\xi_j$ presumably corresponds to the evanescent mode decay comprising of $d$ bands 2 and 3, see Table~\ref{tab:vFermi}, whereas the tail at large $\xi_j$ originates predominantly from the band 6. Overall, one can see that a fit with a single decay exponent, discussed in Eq.~\eqref{eq:J1fit}, is quite oversimplified for a Nb/Ni/Nb junction considered here.

We now turn to the discussion of the effect of exchange splitting energy on the supercurrent in MJJs. So far we have used $M/M_0 = 0.5$, which yields the experimentally observed $\Vex = 0.3$\,eV. It is interesting to investigate how a ferromagnet with a different $\Vex$ (but otherwise the same band structure as Ni) would affect the $w$-dependence of $J_1$. In Fig.~\ref{fig:J1_w_M} we show the results for parametric variations in $M/M_0$. One can see in Fig.~\ref{fig:J1_w_M_orig} that the half-period, $\lambda_J$, and the decay length, $\xi_J$, strongly depend on $M$. Here black points correspond to $M/M_0 = 0.5$, and the self-consistent calculations with no rescaling correspond to $M/M_0=1$. In order to understand how $\lambda_J$ and $\xi_J$ depend on $M$, we perform the fitting procedure Eq.~\eqref{eq:J1fit} for different magnetic moments and plot, in Fig.~\ref{fig:J1_w_M_resc}, the supercurrent density as a function of the rescaled thickness, $(w + \delta_J) / 2{\tilde\lambda}_J$ with ${\tilde\lambda}_J = (M_0/M) \,11.6$\,{\AA}. Remarkably $J_1$, as a function of the rescaled thickness, collapses to the same universal curve. The inset demonstrates that the fitted half-period, $\lambda_J$, is proportional to $1/M$ showing that the oscillation period scales linearly with the inverse of $\Vex$ in this parameter range.%
\footnote{Deviations from linear regime become significant for $M/M_0 \gtrsim 1.7$. For the clarity of the data, we do not show the results for $M/M_0 > 1$ in Fig.~\ref{fig:J1_w_M_resc}.}

We now discuss the difference in crystal orientation in Nb/Ni/Nb junctions. We have performed calculations for Nb/Ni/Nb junctions built from stacking the Ni atomic planes in the (110) orientation instead of the (111) orientation. The results are given in Appendix~\ref{app:Ni110}. Qualitatively, the same physics hold for both stacks built from (111) and (110) Ni planes. However, our calculations show that the actual value for the period of oscillation and for the current decay depend crucially on the details of the electronic structure of the junctions, such as the relative crystal orientation.

Finally, we considered effect of spin-orbit coupling in SFS junctions. Spin-orbit coupling leads to mixing of the minority and majority channels and may change current-phase relationship. The interplay between Zeeman splitting and spin-orbit coupling have been discussed in Ref.~\onlinecite{Cheng:2012}; the regime of interest is Zeeman-field-dominated regime considered there. Indeed, we find that SOC in Ni is much smaller than the exchange splitting because of low atomic number of Ni. As we show in Appendix~\ref{app:spin_orbit_coupling}, the SOC in Ni-based MJJ considered here does not change qualitative picture described above but rather leads to small quantitative changes to the Josephson current.

\section{Conclusion} \label{sec:conclusion}

In this paper we identified two generic mechanisms for the decay of the supercurrent with junction thickness: (i)~exchange-splitting induced gap opening for minority or majority carriers and (ii)~dephasing between different modes due to the significant quasiparticle velocity dispersion with the transverse momentum. It was previously believed that disorder in the ferromagnet is mainly responsible for the supercurrent decay in SFS junctions. In the present work we have shown that band structure effects also contribute to the critical current suppression and thus provide an upper bound for the supercurrent in ideal (i.e. disorder-free) structure.

We found that the Nb/Ni/Nb junction is a suitable system for comparison with the simulations because of the long mean free path in Ni relative to the junction thickness and the quasi-ballistic nature of quasiparticle propagation in the ferromagnet. We have found good agreement with published experimental data for the half-period of the critical current oscillations: $\lambda_J \approx 23$\,{\AA} [see Eq.~\eqref{eq:J1fit} and text after it] versus $\approx 26$\,{\AA} in experiment, Ref.~\onlinecite{Baek:2017}. We have also found that the critical current decays exponentially with the ferromagnet thickness $w$. This is to be contrasted with previously assumed algebraic decay based on results for the clean SFS junctions using simple parabolic-like band structure. We believe that in measured Nb/Ni/Nb junctions with $w \lesssim 50$\,{\AA} the mechanism (i) is likely to be responsible for the supercurrent decay. This finding is crucial for material and geometry optimization of MJJs and superconducting magnetic spin valves.

Understanding the interplay of band structure effects and disorder in MJJs is an interesting open problem. We believe that interfacial disorder due to, for example, surface roughness will mix different $\kpar$ modes and will lead to a larger spread of half-periods. This, in turn, will further enhance the dephasing mechanism (ii) of the supercurrent decay discussed here. Strong disorder in the bulk (i.e. mean free path much smaller than junction thickness $w$) would lead to the diffusive motion of quasiparticles in the ferromagnet which is a significant departure from the quasi-ballistic junction limit considered here. We think that bulk disorder would induce even more dephasing between different modes because phase offsets in this case will depend on different random trajectories of minority and majority carriers. We, therefore, believe that bulk disorder will lead to even stronger decay of the supercurrent with junction thickness, $w$.

\begin{acknowledgments}
HN is much indebted to Dimitar Pashov for stimulating discussions about developing the \texttt{Questaal} package. The authors express their gratitude to Mason Thomas for the organizational help and discussions at the early stages of the project. The authors acknowledge stimulating discussions with Norman Birge, Anna Herr, Tom Ambrose, Nick Rizzo, and Don Miller.

This work is based on support by the U.S. Department of Energy, Office of Science through the Quantum Science Center (QSC), a National Quantum Information Science Research Center. HN and MvS acknowledge financial support from Microsoft Station Q via a sponsor agreement between KCL and Microsoft Quantum. 
In the late stages of this work MvS was supported by the U.S. Department of Energy, Office of Science, Basic Energy Sciences under Award \# FWP ERW7906.
\end{acknowledgments}

\appendix

\section{Mean free path estimate} \label{app:mean_free_path}

In this section we provide an estimate for the mean free path $\lMFP$ in the Ni ferromagnet. Our approach is similar to Ref.~\onlinecite{Gall:2016}. We use the Kubo formula for conductivity
\begin{equation*}
	\sigma_{xx}
	= e^2 \sum_{n,\sigma} \tau_{n\sigma} \, \langle v^2_{n\sigma}\rangle \, {\rho_n(\EF)},
\end{equation*}
where index $n$ labels the Ni bands, $\sigma$ is the spin projection, and $\tau_{n\sigma}$ is the scattering time in each band assumed to be momentum independent. Mean square velocity, $\langle v^2_{n\sigma} \rangle$, and density of states at the Fermi level, $\rho_n(\EF)$, are obtained from our \textit{ab initio} model calculations and are provided in Table~\ref{tab:vFermi}.

We use the available experimental data~\cite{Moreau:2007, Bass:2011} for thick Ni samples $w \gtrsim 50$\,{\AA}: the low temperature linear resistance $1 / (\sigma_{xx}^\uparrow + \sigma_{xx}^\downarrow) \approx 33\,\mathrm{n} \Omega {\cdot} \mathrm{m}$, and bulk spin scattering asymmetry, $\beta_\F = (\sigma_{xx}^\uparrow - \sigma_{xx}^{\downarrow}) / (\sigma_{xx}^\uparrow + \sigma_{xx}^{\downarrow}) = 0.14$. Only a single $6^\uparrow$ band of the majority spin is occupied. Thus, one can readily estimate mean free path for majority carriers $\lMFP^{\uparrow} = \langle v_{\uparrow} \rangle \tau_\uparrow \approx 60$\,{\AA}. For the minority channel we assume that the conductance at large thicknesses is dominated by the most mobile band, i.e. $6^\downarrow$~\cite{Tsymbal:2011}. This gives an estimate for the mean free path $\lMFP^{\downarrow} \approx 61$\,{\AA}, similar to its exchange-split partner $6^\uparrow$. These estimates are consistent with the available ARPES~\cite{Petrovykh:1998} and computational data~\cite{Gall:2016}. As mentioned in the main text, the mean free path for $6^\uparrow$ and $6^\downarrow$ carriers exceeds the typical junction thicknesses measured experimentally.

\section{$\mbox{Nb(110)/Ni(111)/Nb(110)}$ junctions} \label{app:Ni111}

Here we describe how coincident site lattices are constructed for the Nb(110)/Ni(111) interface. Supercells of both Ni and Nb are separately constructed in the following way, to make a coincident site lattice.

The primitive unit cells (\textit{fcc} in the Ni case, with a 0\,K lattice constant 3.515\,{\AA}, and \textit{bcc} in the Nb case with a 0\,K lattice constant 3.295\,{\AA}) are rotated. For Ni, the rotation is compactly described in terms of three Euler angles: rotation about $z$ by $\pi/4$, rotation about $x'$ by $\arccos(\sqrt{1/3})$, rotation about $z''$ by $\arccos(4/\sqrt{19})$. Axes $xyz$ are shown in Fig.~\ref{fig:nbni111nb}; $x'y'z'$ and $x''y''z''$ are axes after first and second rotation, respectively. From the first two rotations the $[1\bar{1}1]$ axis becomes the new $z''$ axis. The last rotation is needed to make it approximately coincident with Nb. The Nb is rotated about $(1,-1,0)$ by $-\pi/2$, about $z'$ by $\pi/4$, and about $z'$ again by $\arccos(5/\sqrt{43})$. After rotations both Ni $(1\bar{1}1)$ and Nb $(\bar{1}\bar{1}0)$ planes are normal to $z$, which is the propagation direction.

Next, superlattices must be constructed. They are generated by scaling the primitive lattice vectors  by the following integer multiples, for Ni and Nb respectively:
\begin{equation*}
	\left[ {\begin{array}{rrr}
		3 & 0 & 2 \\
		-1 & 0 & 4 \\
		1 & 1 & 1
	\end{array}} \right]\!,
	\qquad
	\left[ {\begin{array}{rrr}
		-1 & -4 & 2 \\
		2 & -2 & 0 \\
		1 & 1 & 2
	\end{array}} \right]\!.
\end{equation*}

\begin{figure}[b]
	\centering
	\begin{subfigure}[t]{0\linewidth} \phantomcaption \label{fig:ni110_top} \end{subfigure}%
	\begin{subfigure}[t]{0\linewidth} \phantomcaption \label{fig:ni110_transverse} \end{subfigure}%
	\begin{subfigure}[t]{\linewidth} \includegraphics[width=8.0cm]{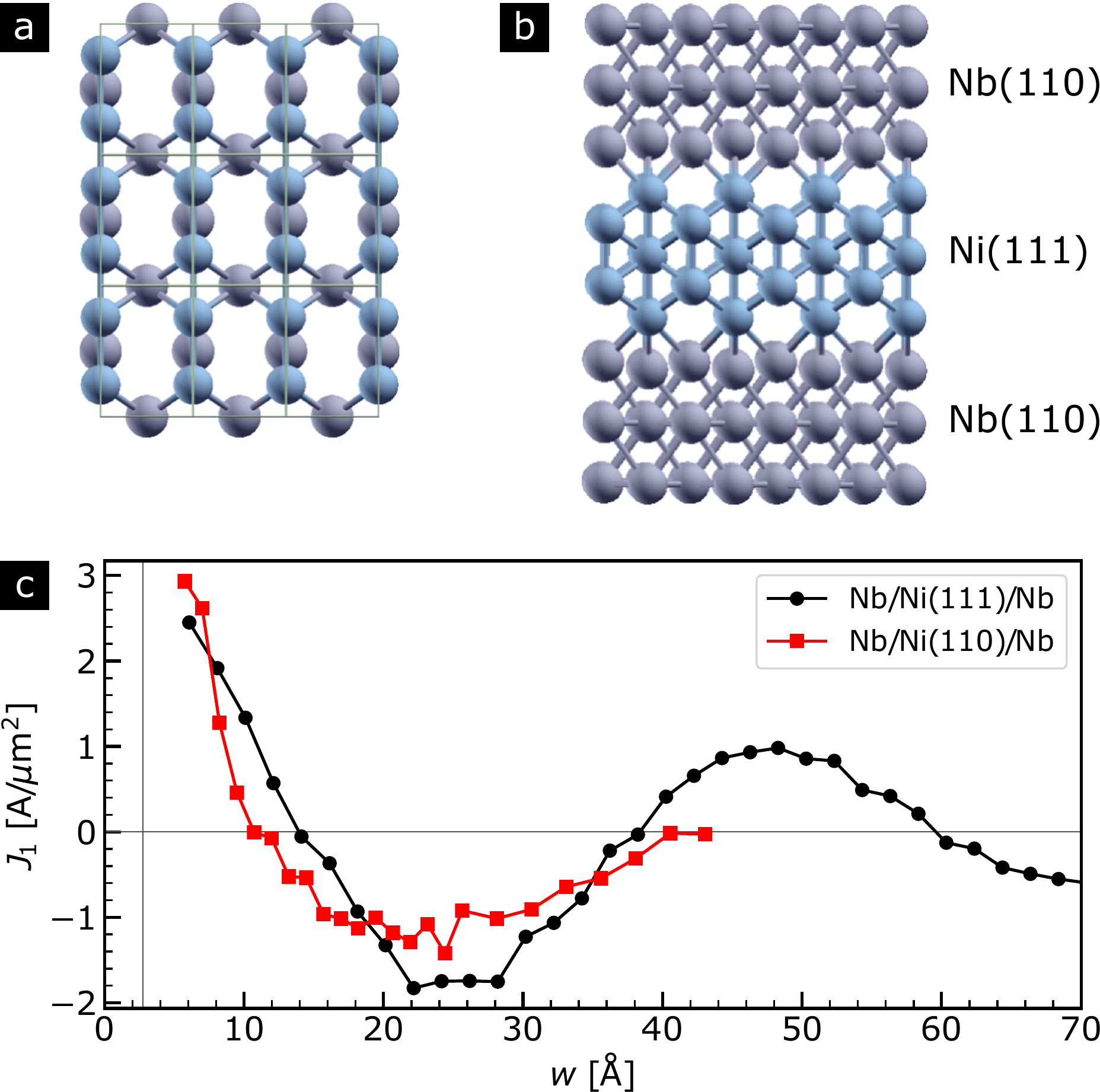} \phantomcaption \label{fig:J1_w_ni110} \end{subfigure}
	\caption{ \label{fig:ni110}
		\subref{fig:ni110_top}~Top view of the Nb(110)/Ni(110) interface supercell. Ni
		(Nb) atoms are shown in light blue (grey). A repetition ($3{\times}3$) of the
		surface supercell is shown. The corresponding Nb(110) and Ni(110) cells contain
		2 atoms each.
		\subref{fig:ni110_transverse}~Transverse view of the Nb(110)/Ni(110)/Nb(110)
		junction with 5 layers of Ni.
		\subref{fig:J1_w_ni110}~$J_1$ component for the Nb(110)/Ni(110)/Nb(110) and the
		Nb(110)/Ni(111)/Nb(110) junctions versus the thickness $w$. Both currents have a
		decaying oscillatory behavior, with a half period of oscillation of $\lambda_J
		\approx 30$\,{\AA} for the Ni(110) case and of $\lambda_J \approx 23$\,{\AA} for
		the Ni(111) case.
	}
\end{figure}

The Ni supercell contains 14 atoms, all in a single plane perpendicular to $z$; the Nb supercell consists of 2 planes with 10 atoms per plane [Fig.~\ref{fig:supercells}]. The lattice vectors transverse to $z$ are nearly coincident, but not identically so. To render them coincident, we opt to shear the Ni by the following linear transformation,
\begin{equation*}
	\left[ {\begin{array}{cc}
		0.997&0 \\
		0.050&1.028
	\end{array}} \right]\!.
\end{equation*}
This shear is a measure of the remaining mismatch of the undistorted Ni and Nb lattices. It is close enough to unity to have a minor effect on the Ni band structure.

Along this axis, Nb planes form an ABAB$\dots$ stacking pattern; the Ni are stacked ABCABC$\dots$ Ni/Nb interfaces are formed by stacking varying numbers of Ni planes on a Nb substrate, and relaxing the supercells to minimize the total energy. This was done for a few small superlattices with 3, 4, and 5 Ni planes, restricting the relaxation to the two Ni and two Nb frontier planes. In this way we can build up structures of arbitrarily many Ni planes, sandwiching unrelaxed planes between frontier planes. Thus the entire set of structures has three families of interfaces: those with integer numbers $3n$, $3n+1$ or $3n+2$ of Ni planes.

Figure~\ref{fig:G} shows that normal conductance, $\G$, experiences $\sim 1.3\%$ deviations from its mean $\langle\G\rangle$ following the pattern of these three families. Overall, both normal conductance and supercurrent (see Fig.~\ref{fig:J1_std_w}) are reasonably smooth functions of the thickness $w$, which suggests that the discreteness of the lattice and the details of lattice relaxation plays a minor role.

\section{$\mbox{Nb(110)/Ni(110)/Nb(110)}$ junctions} \label{app:Ni110}

In this section, we present results for the Nb/Ni/Nb junctions built from stacking the Ni atomic planes with the (110) orientation instead of the (111) orientation. Although the (110) plane orientation is not the most stable configuration for the Nb/Ni interface and most experimental studies focus on (111) orientation, it is interesting to ascertain whether there is an effect of crystallographic orientation on the supercurrent. Indeed, the supercurrent decay and half-period of oscillations with $w$ do depend on crystal orientation.

We study the Nb(110)/Ni(110)/Nb(110) trilayer shown in Fig.~\ref{fig:ni110_top}. The inter-plane distance between the Nb and Ni planes at the interfaces is taken to be the average of the Nb and Ni inter-plane distances. For the sake of simplicity, we do not perform atomic relaxations for this system. The lattice parameter of Nb (3.295\,{\AA}) has been increased to match the lattice parameter of Ni (3.515\,{\AA}) to simplify the construction of periodic supercell. This has a slight effect on the Nb band structure.

Calculations of the trilayers were performed self-consistently, and for transport we reconstruct the potential by rescaling the magnetic moments by 0.5, as in the Ni(111) case. The $0$-$\pi$ and $\pi$-$0$ transitions occur around $w\approx 11$\,{\AA} and 41\,{\AA} [shown in red in Fig.~\ref{fig:J1_w_ni110}], yielding a half-period for the oscillations of $\lambda_J \approx 30$\,{\AA}. This period is significantly larger than $\lambda_J \approx 23$\,{\AA} for the Nb(110)/Ni(111)/Nb(110) case [shown in black in Fig.~\ref{fig:J1_w_ni110}]. The difference of $\approx 7$\,{\AA} between the two half periods represents roughly 5 inter-plane distances for Ni(110) [3 inter-plane distances for Ni(111)], and is not induced by the unrelaxed atomic structures at the Nb(110)/Ni(110) interfaces. It is noteworthy that the calculated period for the Ni(111) trilayers better coincides with available experimental data~\cite{Baek:2017} than for the Ni(110) case.

\section{Effect of spin-orbit interaction} \label{app:spin_orbit_coupling}

\begin{figure}
	\centering
	\begin{subfigure}[t]{\linewidth} \phantomcaption \label{fig:fs_no_soc} \end{subfigure}%
	\begin{subfigure}[t]{\linewidth} \includegraphics[width=8.3cm]{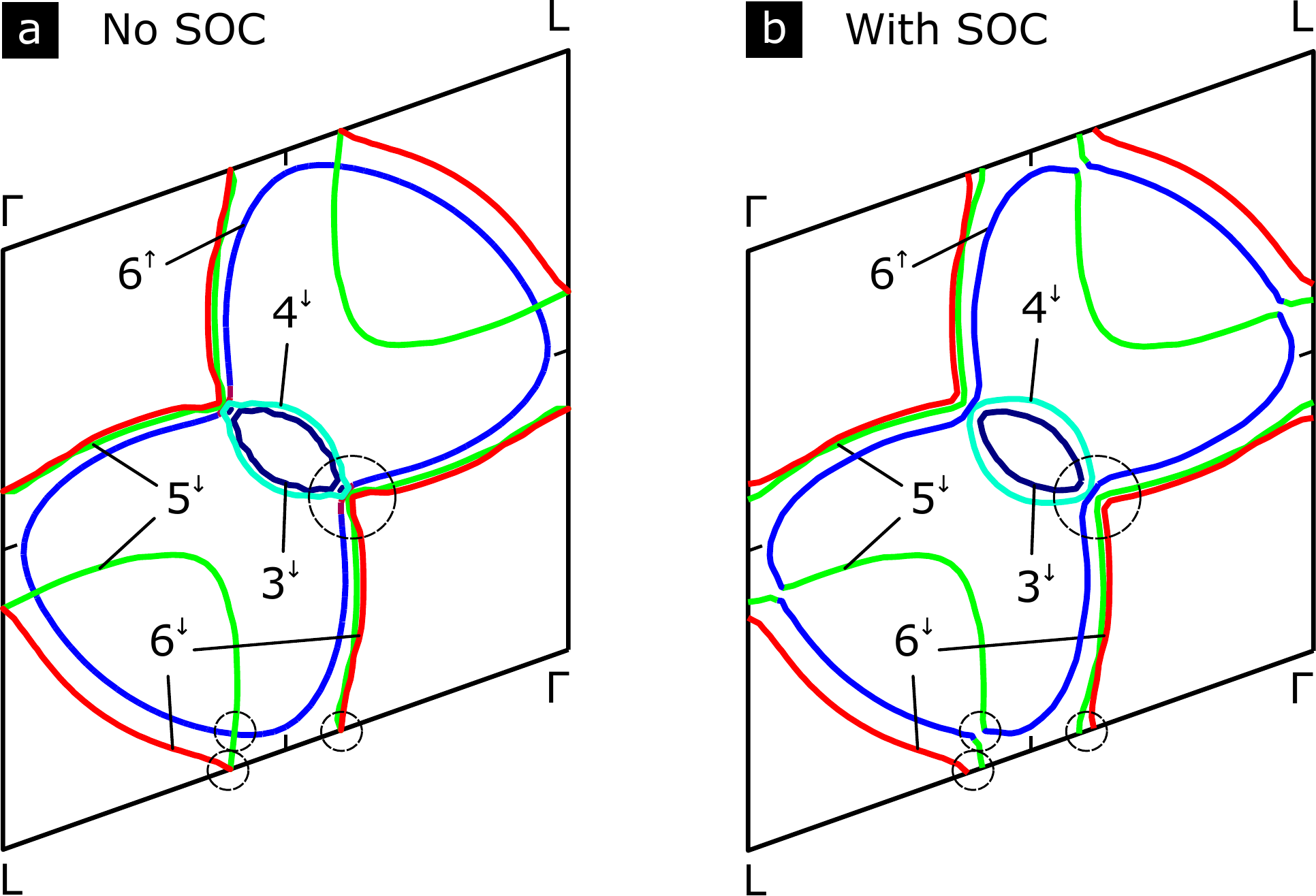} \phantomcaption \label{fig:fs_with_soc} \end{subfigure}
	\caption{ \label{fig:fs_soc}
		\textit{GW} Fermi surfaces of bulk Ni in the $\kpar$ plane corresponding to Ni(111).
		\subref{fig:fs_no_soc}~No SOC [same as Fig.~\ref{fig:fsurfaces}].
		Shown are majority band $6^{\downarrow}$ (red) and 
		minority bands $6^{\uparrow}$ (blue), $3^{\downarrow}$ (dark blue), 
		$4^{\downarrow}$ (cyan), and $5^{\downarrow}$ (green).
		\subref{fig:fs_with_soc}~With SOC.
		SOC opens gaps around the small regions close to the band crossings (circled).
	}
\end{figure}

In this section, we discuss the effect of spin-orbit coupling (SOC) on Josephson current in quasi-ballistic SFS junctions. The effect of spin-orbit interaction is two-fold: (i)~SOC changes the band structure and, therefore, may modify the dependence of the quasiparticle velocities on transverse momenta and (ii)~SOC couples spin and orbital motion and leads to spin precession along the junction. The combination of spin precession and scattering on non-magnetic impurities may introduce random spin-flip processes which suppress the phase difference between minority and majority quasiparticles. We analyze both mechanisms below and show that they represent weak perturbations to our main results and do not change qualitative predictions for quasi-ballistic Ni-based MJJs.

Detailed theoretical analysis of the bulk Ni band structure with SOC and different orientations of the magnetic moment has been discussed in Ref.~\onlinecite{Bunemann:2008}. In Fig.~\ref{fig:fs_soc}, we compare Fermi surface of Ni with and without SOC using the same high-level theory~\cite{Sponza:2017} of Fig.~\ref{fig:bands}. The \texttt{Questaal} code can treat SOC effects, see e.g. tutorial `\href{https://www.questaal.org/docs/numerics/spin_orbit_coupling/}{Spin and spin orbit coupling}.' The SOC implementation is conventional for the energy-dependent local basis set and band structure calculations (see Sec.~2.8.2 in Ref.~\onlinecite{Pashov:2020}), and has also been developed for the Green's functions (see Supplementary Materials in Ref.~\onlinecite{Belashchenko:2015} and Sec.~2.16 in Ref.~\onlinecite{Pashov:2020}).

One may notice that the spin-orbit coupling only weakly modifies the Fermi surface shown in Fig.~\ref{fig:fs_soc}. Indeed, apart from several pockets (close to the degeneracy points) across the Brillouin zone, the SOC only weakly perturbs the band structure. In \textit{fcc} crystals, the $d$-orbitals are split into degenerate $t_\mathrm{2g}$ and $e_\mathrm{g}$ levels, at the $\Gamma$ point, according to the symmetry. The SOC splits further the degenerate $t_\mathrm{2g}$ levels. From these level shifts (taken at the $\Gamma$ $\kpar$-point), we estimate local SOC coupling to be of the order of $\ESO \sim 10\,$meV which is much smaller than the exchange splitting in Ni. In Ref. ~\onlinecite{Bunemann:2008}, the corresponding SOC strength was found to be  68meV which further corroborates our conclusions.         

One can now evaluate effect of SOC coupling on quasiparticle propagation through the junction. Given that bulk band structure of Ni is centro-symmetric, Dresselhaus spin orbit coupling is forbidden by symmetry. The Rashba SOC may appear due to the inversion symmetry-breaking along the direction of the junction. However, the rapid screening of the interface potential in the ferromagnet would limit Rashba SOC to a few atomic layers in the junction. Finally, we believe that effect of local spin-flip processes due to scattering on impurities is weak in the regime of interest. Indeed, here we are considering quasi-ballistic magnetic Josephson junctions (i.e. the thickness of the junction smaller than the mean free path $\lMFP$ in Ni). Thus, we expect impurity-induced spin-relaxation rate to be suppressed as well.

\section{Normal scattering matrix calculation} \label{app:scattering_matrix_calculation}

The \texttt{Questaal} code is based on the LMTO technique which uses a set of electron wave-function $\phi_{\mathrm{R}L}$ and its energy-derivative $\dot\phi_{\mathrm{R}L}$ as a basis set.  The wave-functions $\phi_{\mathrm{R}L}$ are solutions of the spherical Schr\"odinger equation in a sphere around a given atom at position $R$ with angular quantum numbers $L = l, m$. The LMTO technique is an all-electron approach, with does not rely on the use of pseudo-potential. The basis set contains core electrons as well as `valence' (non-core) electrons.

The electron ground state of the systems is obtained with the DFT-LDA-ASA framework~\cite{Pashov:2020}. The (energy-dependent) basis set consists of partial waves of \textit{s}, \textit{p}, \textit{d} character on each of the atomic sites.  Core levels are integrated separately from the valence partial waves to obtain the all-electron charge density.  However, they are not included in the secular matrix.

The calculations are converged when variations in the electron density and total energy between the last iterations, are below $3\times 10^{-5}$ ($10^{-5}$) respectively.  To obtain the self-consistent charge density, calculations were performed with a $\kpar$ mesh of ($8 \times 4$).  As is well known, for purposes of determining the density, a finer $\kpar$ mesh is not needed since the output is a $\kpar$-independent potential for a subsequent transport calculation. 

Atomic relaxations are performed on relatively small cells with a full-potential method~\cite{Pashov:2020}, with periodic boundary conditions.  Once the ground-state is reached, the atomic positions of the frontier atoms at the Nb/Ni and Ni/Nb interfaces are allowed to relax to minimize the total energy. Convergence is achieved when all the forces on the relaxed atoms are below $\sim 25$\,mRy/bohr.  Atomic relaxations are constrained to the first and second frontier layers; these shifts are then added to the ideal geometry of the (larger) stacked cells in the subsequent ASA calculations of transport.  More details of the calculations can be found in the \texttt{Questaal} tutorial `\href{https://www.questaal.org/tutorial/application/nbni_superlattice/} {Nb/Ni superlattice}.'

A denser $\kpar$ mesh is needed as the transmission (reflection) probability is a quantity very sensitive to the $\kpar$-point sampling. As shown in Fig.~\ref{fig:J1_G_w} of the main text, we used a mesh of ($24 \times 24$) and a mesh of ($90 \times 92$) points, giving rise to 290 and 4142 irreducible $\kpar$ points respectively.  This establishes that our transport calculations are well converged in $\kpar$.

Note that the principal layer technique (used for the transport) does not rely on periodic boundary conditions in the (transport) direction perpendicular to the layers (in our case the $z$-axis).  Hence there is not $k_z$ point sampling, the wave number $k_z(E)$ should be understood as a continuous complex function of $E$.

Further details for the constructions of the family of the $3n$, $3n+1$ or $3n+2$ of Ni planes of the Nb(110)/Ni(111)/Nb(110) junctions can be found in the \texttt{Questaal} tutorial `\href{https://www.questaal.org/tutorial/lmpg/lmpg_nbni/} {Nb(110)/Ni/Nb(110) metallic trilayers}.'

\bibliography{mjj}

\end{document}